\documentclass[aps, twocolumn, nopacs, amsmath]{revtex4-1}

\usepackage{mathtools, amsmath, amssymb, graphicx, amsfonts}
\usepackage{algorithm}
\usepackage{algorithmic}

\usepackage{xcolor, comment} 
\usepackage{bbm} 
\usepackage{float} 
\usepackage{amsopn} 
\usepackage{mathrsfs} 

\newcommand{\im}{\operatorname{im}}

\newcommand{\nullity}{\operatorname{null}}

\newlength\myindent
\def\bea{\begin{eqnarray}}
\def\eea{\end{eqnarray}}

\begin{document}

\title{Spectral Detection of Simplicial Communities via Hodge Laplacians}

\author{Sanjukta Krishnagopal}
\email{s.krishnagopal@ucl.ac.uk}
\affiliation{Gatsby Computational Neuroscience Unit, University College London, London,  WC1E 6BT, United Kingdom }

\author{Ginestra Bianconi}
\email{ginestra.bianconi@gmail.com}
\affiliation{School of Mathematical Sciences, Queen Mary University of London, London, E1 4NS, United Kingdom}
\affiliation{The Alan Turing Institute, 96 Euston Road, London,  NW1 2DB, United Kingdom}


\begin{abstract}

While the study of graphs has been very popular, simplicial complexes are relatively new in the network science community. Despite being are a source of rich information, graphs are limited to pairwise interactions. However, several real world networks such as social networks, neuronal networks etc. involve simultaneous interactions between more than two nodes. Simplicial complexes provide a powerful mathematical way to model such interactions. 
Now, the spectrum of the graph Laplacian is known to be indicative of community structure, with nonzero eigenvectors encoding the identity of communities. Here, we propose that the spectrum of the Hodge Laplacian, a higher-order Laplacian applied to simplicial complexes, encodes simplicial communities. 
We formulate an algorithm to extract simplicial communities (of arbitrary dimension). We apply this algorithm on simplicial complex benchmarks and on real data including social networks and language-networks, where higher-order relationships are intrinsic. Additionally, datasets for simplicial complexes are scarce. Hence, we introduce a method of optimally generating a simplicial complex from its network backbone through estimating the \textit{true} higher-order relationships when its community structure is known. We do so by using the adjusted mutual information to identify the configuration that best matches the expected data partition. Lastly, we demonstrate an example of persistent simplicial communities inspired by the field of persistence homology.
\end{abstract}

\maketitle
\section{Introduction}

The increasingly popular interdisciplinary field of network science \cite{BAR13} aims to capture properties of systems through their interactions. Interactions are ubiquitous in nature, and applications of network science range from firing neurons in the brain \cite{TEL11}, the dynamics of social interactions \cite{HOG08}, biological systems \cite{GOS18}, climate, transportation networks \cite{HAV12}, or  the stock market \cite{KIM17}.

Network approaches are very successful at extracting the rich interplay between structure and dynamics \cite{BAR08}.  Conventionally, a network captures the interactions between two nodes (or 'vertices') as the properties of the link (or the `edge') connecting them. However it has been realised that pairwise networks describing a single type of interaction can be too restrictive to model the set of interactions existing in a complex system.
The need for modeling multiple types of interactions has led to innovation in multi-layer networks, where different layers represent different types of interactions \cite{BIA18}.  Mounting evidence suggests that another limitation of networks resides in the pairwise nature of their interactions. Indeed  the rich set of  interactions between the elements of a complex system are better captured through a model that allows for simultaneous interactions between more than two entities \cite{bassett,bianconi2021higher,battiston,tina}.  For example, consider a social network modeling the interaction of students in a university during lunch break. Here groups of 3 or more students emerge just as naturally as groups of 2. Modeling a group of 3 individuals as 3 sets of pairwise interactions is fundamentally different, and misleading, compared to simultaneously modeling the interaction of all three. Such a higher-order interaction can, for instance, be represented as a filled triangle, differentiating it from a set of three edges. Indeed, simplicial representations involving filled triangles, tetrahedra and higher dimensional simplices have provided cohesive explanations for complex dynamics in neuroscience \cite{SIZ18, AND20}, protein interaction \cite{EST18}, complex systems \cite{SAL18,BEN16}, signal processing \cite{SCH20}, disease spreading \cite{IAC19} etc. With this in mind, graph-structures such as simplicial complexes and hypergraphs  \cite{EST05} are gaining large traction in recent years. 
\\
Higher-order interactions can be captured by simplicial complexes as well as by hypergraphs. 
The difference between them is subtle, simplicial complexes are closed under the inclusion of subsets, while hypergraphs are not. Hence, simplicial complexes are topological spaces, and lend themselves to analysis from the lens of  topology, a rich and heavily researched field of mathematics. The tools from simplicial topology can be exploited in the analysis of networks and simplicial complexes.

Exploiting the relationship with topology, recent works have investigated higher-order dynamics \cite{joaquin,millan,reza,calmon} and data analyses using Hodge theory \cite{JIA11}, as well as persistent homology \cite{STO17}. Additionally, applied topology studies the underlying properties such as the Betti numbers (the number of high-dimensional holes) of simplicial complexes applied to real data. Topology is also fundamental for proposing models of emergent geometry and for revealing the interplay between topology and synchronization dynamics \cite{WU15,BIA16,BIA17}. 
\\
The advent of community detection in conventional graphs \cite{GIR02,FOR16, BLO08, SHA17} has had a significant impact in the understanding of complex networks. It can offer insight into how edges are organized within the network, and guide dynamics on the network. Nodes that have many edges (or edges with high weight) between them tend to belong to the same community, whereas nodes that have few edges (or edges with low weight) between them tend to fall in different communities. 
Among the most successful algorithms for community detection the clique community detection \cite{palla2005uncovering} was the first to propose an algorithm allowing for overlapping communities. The $k$-clique community is indeed an algorithm that partition the $k$-cliques (fully connected subgraphs of $k$ nodes) of a network into communities where the $k$-cliques of a given community are connected by a path alternating $k$-cliques and $(k-1)$-cliques formed by a subset of their nodes.
Community detection methods are a rich source of information and have been widely applied to extract patterns of interactions in brain networks \cite{BET20,VAN11}, epidemiology \cite{DAS20}, power-grids \cite{CHE15}, opinion dynamics \cite{MOR10} etc.
\\
Naturally, the analog of community detection on hypergraphs simplicial complexes can provide important insight into their structure and dynamics \cite{BEN16,EBL19,DER05,BIL19,chodrow2021hypergraph,eriksson2021flow,carletti2021random}. While spectral community detection \cite{capocci2005detecting,von2007tutorial} in graphs in conventional graphs is rather well studied, there exists surprisingly limited work on spectral community detection in simplicial complexes.  
In this work, we study  simplicial communities that generalize and extend clique communities to general simplicial complexes and we reveal the relationship between simplicial communities and the spectrum of the higher-order Laplacian. 

The eigenvalue spectrum of the graph Laplacian is known to encode several properties of the graph itself. For instance, the number of communities is given by the dimension of the kernel of the Laplacian. Additionally, the zero eigenvectors take constant values on communities. Spectral community detection methods exploit community structure captured by the sign of the eigenvectors associated with small eigenvalues \cite{NEW13}. Here we propose that simplicial communities can be encoded in the spectrum of a higher-order Laplacian also known as the Hodge Laplacian. In other words, the eigenvectors of the higher-order Laplacian have support (the simplices on which the vector takes non-trivial values) localized on simplicial communities. 
Communities of higher dimensional simplices in a simplicial complex can be defined in two ways - through connections in higher-order order simplices and lower order simplices. For instance, one can identify communities of triangles that are connected by sharing edges (down/ lower-dimensional simplices), or by being faces of the same tetrahedra (up/ higher-dimensional simplices). The Hodge Laplacian can be decomposed into the down and up Laplacians that yield down and up communities respectively. The down communities consist of $k$-simplices that are $k-1$connected, which is analogous to the concept of clique-communities introduced in \cite{palla2005uncovering}. The $k$-up communities are isomorphic to the $(k+1)$-down communities. Hence, $k$-up communities are identifiers of $(k+1)$-clique-communities. 
\\
Topological spaces such as simplicial complexes lend themselves very well to topology theory. Deriving from Hodge decomposition,  any chain of simplicial complexes can be decomposed into a sum of three independent spaces: up-communities, down-communities and harmonic representatives, the last of which is known to correspond to topological holes. Here we discuss the implications of this decomposition and the relationship with topology. Additionally, we validate our approach on several synthetic simplicial complexes of various styles, including ones with and without topological holes. Lastly, we implement simplicial community detection on three real datasets - the famous Zachary karate club network \cite{ZAC77}, a network-science collaboration network, and language networks denoting word-interactions in the book `Les Mis\'erables' by Victor Hugo. It is worth noting that language networks naturally contain layered structure that can be represented as high-dimensional simplices.
Our analysis of the Zachary karate club deserves a particular mention because this application highlights the important difference between the simplicial communities and the clique communities. In fact if we start from a network, clique community detection makes the assumption that all cliques are filled, i.e., it assumes that each clique indicates a  many-body interaction which might not be realistic. Indeed, since simplicial datasets are uncommon, the inference of true many-body interactions starting from the exclusive knowledge of pairwise interactions (a network) has been receiving increasing attention \cite{young2021hypergraph,musciotto2021detecting}.
Here we show, by analysing the Zachary Karate club network, that  given the ground-truth community of a network, the study of simplicial communities can be turned to  an inference algorithm for determining the true higher-order interactions between the nodes of a given network.
Therefore this case study highlights the importance of distinguishing between true simplicial communities and the clique communities of a network.

The paper is organized a follows. In Sec II we define graphs, simplicial complexes and clique complexes; in Sec III we define simplicial communities and clique communities and we highlight their similarities and differences; in Sec IV we summarize the main spectral properties of graphs and simplicial complexes and we emphasize the important role of the Hodge decomposition and its physical interpretation; in Sec. V we reveal the relation between simplicial communities and the spectral properties of simplicial complexes; in Sec VI we formulate a spectral clustering able to detect simplicial communities; in Sec VII we apply this algorithm to simplicial complex benchmarks; in Sec VIII we study real network data by inferring and extracting their simplicial communities; finally in Sec. IX we provide the concluding remarks. The paper is enriched by two Appendices providing the necessary background in algebraic topology and providing additional information about the identity of the obtained simplicial communities of the real networks analysed in this work.
\section{Networks, simplicial complexes and clique complexes}
\subsection{Graphs}

An undirected graph $G=(N,E)$ consists of a set of vertices $N$ and a set of edges $E$ that represents elements of a system and their interactions respectively. Examples of networks are the World-Wide-Web, Facebook, ecological networks, brain networks etc. The structure of an unweighted graph can be encoded in its adjacency matrix ${A}$ of elements $A_{ij}= 1$, if node $i$ is connected to node $j$ via a link or an edge, and $A_{ij}= 0$ otherwise. In weighted graphs, the adjacency matrix takes on values of the edge weights.  

\subsection{Simplicial Complexes}
\begin{figure}[htbp!]
\includegraphics[width=0.9\columnwidth]{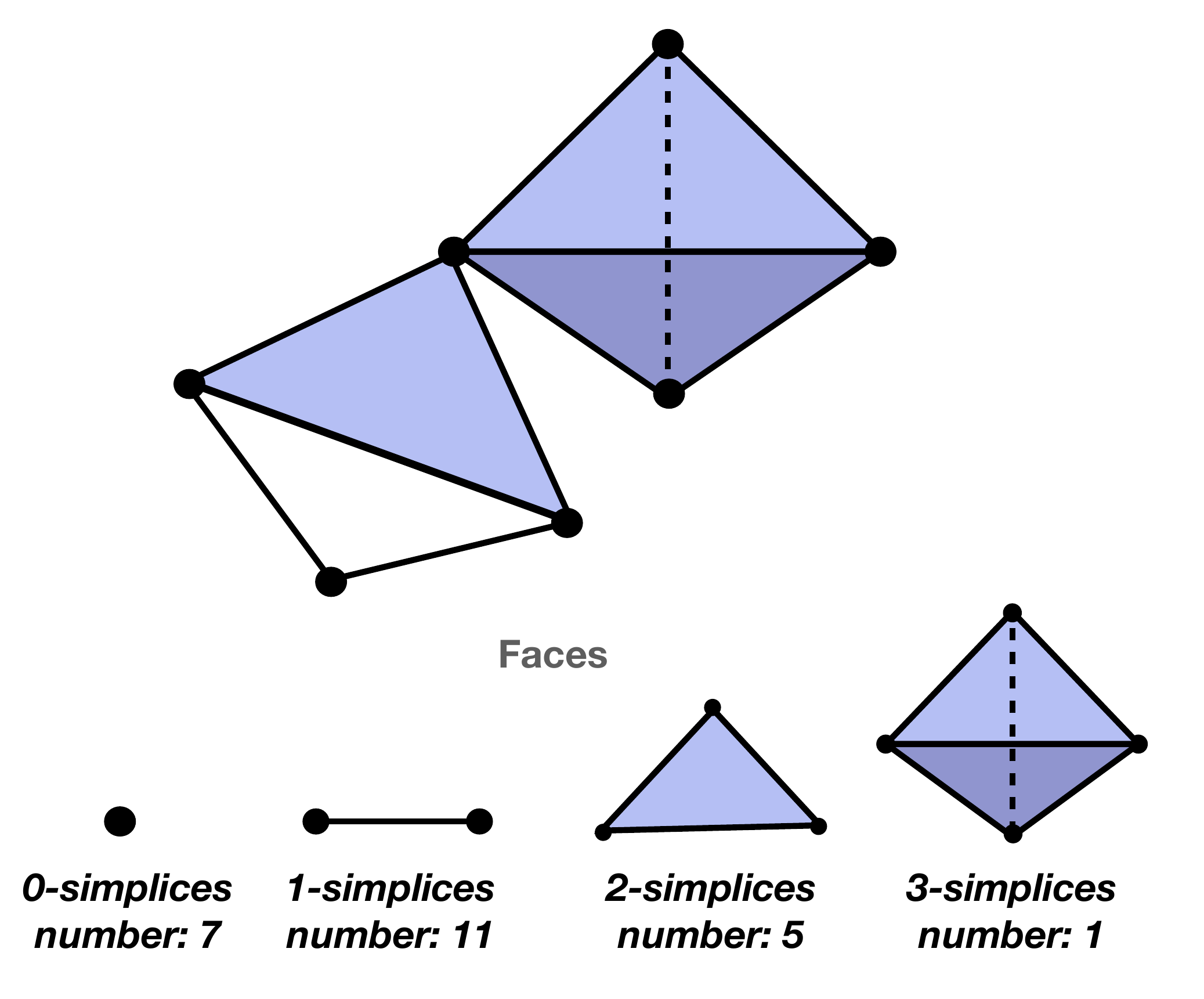}
\caption{A simplicial complex and its decomposition in d-dimensional simplices. The number of $k$-simplices in the top simplicial complex are listed.}
	\label{faces}
\end{figure}

Graphs are unable to capture multi-node interactions which are fundamental in modeling several systems. These can be explained by a mathematical framework called {\em simplicial complexes}, which is a higher-order network.
For instance, in a network, three individuals that wrote a paper together would be denoted by a triangle with three edges indicating three pair-wise interactions. However, in topology, this is denoted by a filled triangle (also known as a 2-simplex) indicating a simultaneous 3-way interaction. Specifically, given a set of $l$ nodes ${n_0, n_1 \ldots, n_l} \in N$ in a network, a $p-$simplex is a subset $\sigma_p = [{n_0,n_1,\ldots, n_p}]$ of $p$ nodes and a $q-$face of  $\sigma_p$ is a set of $q$ nodes (for $q<p$) that is a subset of the nodes of  $\sigma_p $.
A \emph{simplicial complex} ${K}$ consists of a set of simplices, that are closed under inclusion:
\bea
\tau\subseteq\sigma \Rightarrow \tau \in {K} \text{~for any}\ \sigma\in {K},
\eea
where `$\subseteq$' denotes the subset relation between $\sigma$ and $\tau$, two subsets of the simplicial complex. When $\tau\subseteq \sigma$, we say that $\tau$ is a \emph{face} of $\sigma$, which by the inclusion axiom implies \emph{every face of a simplex is again a simplex}. Fig.~\ref{faces} shows examples of faces of a simplicial complex. 

We use $\vert \sigma\vert$ to denote the dimension of a simplex $\sigma$. The \emph{dimension} of a simplex equals the number of vertices in the simplex minus one; for instance $0$-dimensional simplices are nodes and $1$-dimensional simplices are edges. The dimension of a simplicial complex is the largest dimension of its simplices. By ${S}_k$ we will denote the  set of $k$-dimensional simplices, i.e. as
\bea
{S}_k := \lbrace \sigma\in S : \vert \sigma\vert = k+1\rbrace,
\eea

We call the simplices in ${S}_k$ the $k$-simplices of ${K}$ and let $\Gamma_{[k]}$ denote the number of $k$-simplices in the simplicial complex.
Interestingly it is possible to reduce a simplicial complex to a network called {\em the simplicial complex skeleton} by retaining only the nodes and the edges of a simplicial complex.

Note that there is a natural correspondence between hypergraphs, that is a rapidly growing topic of study in networks, and simplicial complex (a facet of a simplicial complex corresponds to an edge in a hypergraph). However, simplicial complexes also allow the use of powerful mathematical tools from topology  that aren't directly applicable to general  hypergraphs. Despite this, some progress has been made in this direction for oriented hypergraphs \cite{jost2019hypergraph,mulas2020coupled}.

\subsection{Clique Complex and Network Skeleton}
A $k$-clique of a network is  a fully connected subgraph of the network including exactly $k$ vertices. 
A clique complex $\Delta(G)$ \cite{bianconi2021higher,kahle2009topology} of an undirected graph $G$ is a simplicial complex in which each $k$-clique of the network is considered a $(k-1)$-dimensional simplex of the simplicial complex.  For instance a $3$-clique of the network $G$ is treated as a $2$-simplex of the clique complex $\Delta(G)$. Since a subset of a clique is itself a clique, the clique complex is closed under the inclusion of the faces of every simplex belonging to it. In other words, the clique complex fills every triangle, tetrahedra and higher-order structures to form simplices.

Therefore the clique complex of a network provides a way of generating simplices that describe the network topology without having detailed information about the existence of specific many-body interactions captured by the simplices, it simply fills all triangles, tetrahedra and all higher-dimensional structures.
Interestingly scale-free networks have a rich clique structure and are known to have a diverging clique number even in the sparse regime \cite{clique}. This implies that their clique complex can have a large dimension even if the original network does not explicitly includes many-body interactions. A similar phenomenon can be encountered by starting from the configuration model of simplicial complexes \cite{courtney, YOU17}, and by generating the clique complex from their network skeleton. Interestingly enough this set of operations in general will not produce the original simplicial complex as the clique complex can contain more simplices than the original simplicial complex.

From a Data-Science perspective a crucial inference problem involves the extraction of true many-body interactions from pairwise network data. For instance in a scientific collaboration network this would entail predicting teams of two or more co-authors only on the basis of the pairwise information of who collaborated with whom.
The clique complex of a network provides the simplicial complex with maximum number of simplices compatible with  original network (i.e. whose network skeleton is the original network) through filling all higher-order structures. However the true many-body interactions may be captured by a simplicial complex including only a subset of the simplices present in the clique complex (e.g. filling only a subset of all triangles), hence the need to formulate reliable inference methods to detect which cliques of the network correspond to filled simplices such as the ones proposed in \cite{young2021hypergraph,musciotto2021detecting}. As such, a simplicial complex is a subset of a clique complex.

\section{ Simplicial Communities and Clique Communities}

\subsection{Simplicial Communities}

Two simplices $\sigma$ and $\hat \sigma$ are $k$-connected if there is a sequence of simplices $\sigma, \sigma_1, \sigma_2 \ldots, \sigma_n, \hat\sigma$ such that any two consecutive simplices share at least one $k$-face (a simplex with $k+1$nodes). For instance, a set of triangles ordered such that consecutive pairs of triangles share a node would be considered $0$-connected, and if consecutive triangles shared an overlapping edge, they would also be considered $1$-connected. A simplicial complex is considered to be $k$-connected if any two simplices of dimension greater than or equal to $k$ are $k$-connected.
In the simplest case, the network skeleton of a $0$-connected simplicial complex is a connected graph.
If a simplicial complex is not $k$-connected, the simplicial complex contains more than one $k$-simplicial community.

Consider a simplicial community partition the $k$-simplices of the simplicial complex into $c_k$ {\em $k$-up communities} $\{\pi_1 , \ldots, \pi_{c_k}\}$. Each $k$-up community is formed by  a set of $k$-simplices that are $(k+1)$-connected, i.e.,  there exists an ordered chain of $k$-simplices such that consecutive simplices are faces of the same $(k+1)$-simplex. Let us denote the $k$-simplicies in the simplicial communities  by $\pi_1 , \ldots, \pi_{c_k}$ where $\pi_i$ is the set of all $k$-simplices in the $i^{th}$-simplicial  community, with
$\pi_i \cap \pi_l=\emptyset$ for $l\neq i$.
The induced partition on $(k+1)$-simplices is denoted by  $\Pi_0, \Pi_1 , \ldots, \Pi_{c_k}$ where $\Pi_i$ is the set of all $(k+1)-$simplices in the $i^{th}$ $(k+1)$-down community (or $(k+1)$-clique community), i.e., community of $(k+1)$-simplices. Each $(k+1)$-down community is formed by  a set of $(k+1)$-simplices that are $k$-connected, i.e., there exists an ordered chain of $(k+1)$-simplices such that consecutive simplices have an overlapping $k$-face.

It follows that  the $k$-up community is isomorphic to the $(k+1)$-down community, i.e., the $k$-simplices are simply the corresponding faces of the $(k+1)$-down communities. In the rest of this paper, the absence of a directional specifier (up/down) (e.g. $k$-simplicial community) refers to the $k$-up simplicial community.

\subsection{Clique Communities}
Two $k$-cliques that share a common $(k-1)$-clique, are considered to be lower adjacent. For instance, two triangles with a common edge are lower adjacent. A $k$-clique community of a graph $G$ can be defined as a set of $k$-cliques such that there there exists a sequence of  adjacent $k$-cliques between any two $k$-cliques within the community. In order words, a $k$-clique community is a maximal union of $k$-cliques that are pairwise connected, analogous to connected components in graphs. Clique communities have served as an effective tool in analyzing properties of a network such as community structure and higher-order connectivity. Important applications of this include in biology, economics, social dynamics etc. The first approach for computing clique communities for a \textit{network} was introduced in \cite{PAL05} that uses the the Bron—Kerbosch algorithm. Here, all maximal cliques in a network are identified, and then clique communities are extracted using a clique-clique overlap matrix. Since then, various extensions of this approach have been proposed \cite{FU14,HAO15,GRE12}. 

Interesting, the $(k+1)$-clique communities of a network reduce to the $k$-down simplicial communities of its clique complex.
However the true $k$-down simplicial communities of the simplicial complex including only the true many-body interactions existing between the set of nodes of the original network can differ significantly from the $(k+1)$-clique communities of the network. 

\section{Spectral Properties of Networks and Simplicial Complexes}

\subsection{Graph Laplacian}

The graph Laplacian \cite{chung1997spectral} is an operator that describes diffusion on a network and has profound effect on synchronization dynamics. As such the graph Laplacian is crucial to understand the relationship between network structure and dynamics.

The graph Laplacian matrix is defined as \bea{ L}_{[0]}={ D} -{ A},\eea where ${ D}$ is a diagonal matrix whose elements are the degrees of the nodes and $A$ is the adjacency matrix of the network.
The graph Laplacian ${ L}_{[0]}$ can be written in terms of this boundary operator as follows: 
\bea
{ L}_{[0]} = B_1 B_1^T,
\label{LB1}
\eea
 where  the boundary operator $B_1$ is a map from edges to nodes:
\bea
B_1(i,\ell)=\left\{\begin{array}{ll}
	-1&\mbox{if} \ \ell=[i,j],\\
	1 &\mbox{if} \  \ell=[j,i],\\
	0 & \mbox{otherwise}.\end{array}\right.
\eea 
for a node $i$ and a edge $\ell$.
The expression given by Eq. (\ref{LB1}) show very explicitly that the graph Laplacian is a positive semi-definite operator, whose eigenvalues are non-negative.
The spectral properties of the graph Laplacian encode important information about the topology of the graph \cite{chung1997spectral}.
 In particular  the degeneracy of the smallest (zero) eigenvalue of the graph Laplacian  corresponds to the number of connected components in the graph. Additionally, the spectral gap  of the graph (also known as Fiedler's gap) is the smallest non-zero eigenvalue of ${L}_{[0]}$ and is indicative of how `separated' the two graph communities are. Indeed several \textit{spectral clustering} methods exploit this property for community detection \cite{NEW13}. In this framework graphs communities are  obtained from the sign of the components of the corresponding Fiedler eigenvector.
 
.


%
\subsection{The Hodge Laplacian}
The topology of simplicial complexes can be investigated with the powerful tool of algebraic topology. Algebraic topology allows the generalization of the graph Laplacian to higher-order Laplacians, also called Hodge Laplacians, \cite{horak2013spectra,bianconi2021higher} which describe higher-order diffusion and carry important topological information about the simplicial complex on which they are defined.

In algebraic topology, each simplex of the simplicial complex is assigned an orientation, where one can show that the choice of the ordering does not affect the spectral properties of the Hodge Laplacians as long as the orientation is induced by the nodes labels. For instance  one can assign  a positive orientation to the simplicies whose vertices are listed according to a  positive ordering of their labels and a negative orientation to simplicies whose vertices are listed according to a negative ordering of their labels
 (see Appendix \ref{app:orientation} for more details).
 
On a simplicial complex one can define the  $k^{th}$-boundary operator  as linear map from oriented $k$-simplices to the oriented $(k-1)$-simplices in their boundary.
The $k^{th}$ boundary operator $\partial_k$ can be represented by a $m \times n$ matrix $B_k$  where  $m$ is the number of $(k - 1)$-simplices and $n$ is the number of $k$-simplices of the simplicial complex (see \ref{app:boundary} for definition).
The $k^\text{th}$ higher-order Laplacian $L_k$, also called the  \emph{Hodge Laplacian} \cite{LIM20}, for $k>0$ is defined as follows
\begin{align}
L_k = L_k^{down}+ L_k^{up}  =  &B_k^T B_k + B_{k+1} B_{k+1}^T.
\label{eq:Hodgelaplacian}
\end{align}
where 
\bea
L_k^{down} &=& B_k^T B_k,\nonumber \\
L_k^{up} &=& B_{k+1} B_{k+1}^T.
\label{Lup}
\eea For $k=0$ the Hodge Laplacian is simply the graph Laplacian of the network skeleton of the simplicial complex, i.e.
\bea
L_{[0]} = L_0^{up}= B_1 B_1^T.
\eea
 The higher-order up and down Laplacians have matrix elements given by 
\bea
	L_k^{up}({\sigma,\hat{\sigma}})=\left\{\begin{array}{ll}
	d_k^{u}(\sigma)&\mbox{if} \ \sigma=\hat\sigma,\\
	-1 &\mbox{if }\Omega ( \sigma) =  \Omega ( \hat{\sigma}),\\
	1&\mbox{if} \ \Omega ( \sigma) = - \Omega ( \hat{\sigma}),\\
	0& \mbox{otherwise}.
	\end{array}\right. 
	\label{Lupe}
	\eea
	\bea
	L_k^{down}({\sigma,\hat{\sigma}})=\left\{\begin{array}{ll}
	k+1&\mbox{if} \ \sigma=\hat\sigma,\\
	-1 &\mbox{if} \  \Omega ( \sigma) =  \Omega ( \hat{\sigma}),\\
	1&\mbox{if} \ \Omega ( \sigma) =  -\Omega ( \hat{\sigma}),\\
	0&\mbox{otherwise}.
	\end{array}\right. 
	\label{Ldowne}
	\eea
where  $d_u^k(\sigma)$ is number of $k+1$ simplices having  $\sigma$ as one of their faces, and where $\Omega(\cdot)$ is the orientation of the simplex within the parenthesis. 
The up and down Laplacians can also be proven to be independent on the orientation of the simplices if the assigned orientation of the simplices is induced by a labelling of the nodes. We will denote $up/ down$ by superscripts $u/d$ respectively. 

The main property of the Hodge Laplacian used by topologists is that the degeneracy of the zero eigenvalue of $L_k$ is equal to the Betti number $\beta_k$ and that their corresponding eigenvectors localize around the corresponding $k$-dimensional cavity of the simplicial complex. Therefore Hodge Laplacians with $k > 0$ are \textit{not guaranteed to have a zero eigenvalue}, unlike graph Laplacians. 

\subsection{Hodge Decomposition}
From the definition of the Hodge Laplacian it follows that the Hodge Laplacian is real, symmetric and \emph{positive semidefinite}.
Interestingly the $k^{th}$-up Laplacian, the $k^{th}$-down Laplacian,  and their sum $L_k$ commute with each other and can be simultaneously diagonalized. Moreover we have 
\bea
\mbox{im}(L_k^{down})&\subseteq&\mbox{ker}(L_k^{up}),\nonumber\\
\mbox{im}(L_k^{up})&\subseteq&\mbox{ker}(L_k^{down}),\nonumber \\
\mbox{ker}(L_k)&=&\mbox{ker}(L_k^{up})\cap\mbox{ker}(L_k^{down}).
\label{imker}
\eea
Therefore an eigenvector of $L_k$ corresponding to a non-zero eigenvalue $\lambda$ is either a non zero eigenvector of $L_k^{down}$ or a non-zero eigenvalue of $L_k^{up}$.
This is a central result of Hodge theory called {\em Hodge decomposition} which can be used to decompose the space on which $L_k,L_k^{up}$ and $L_k^{down}$ act. This is  the space  $C_k$  of all $k$-chains, i.e. the set of all linear combinations of the $k$-simplices of the simplicial complex (see Appendix \ref{app:boundary} for detail).
In particular the Hodge decomposition can be stated as: 
\begin{equation}
C_k=\mbox{im}(B_k^{\top})\oplus\mbox{ker}(L_k)\oplus\mbox{im}(B_{k+1}).
\label{eq:hodgedecomposition}
\end{equation}
For $k=1$, this expression indicates that  any $1-$chain can be decomposed into the sum of three orthogonal elements: a gradient (in the image of $B_1$), a curl (in the image of of $B_{2}^T$), and a harmonic representative (in the kernel of $L_1$). There exists an analog for arbitrary $k$ where one can conceive a higher-dimensional curl and gradient operator.
The Hodge decomposition has played an important role in several analyzes and applications. For instance Hodge decomposition is central for defining higher-order synchronization of $k$-chains and of coupled chains of different dimension \cite{millan,joaquin,reza,calmon}. Moreover, the space of $1$-chains on simplicial complexes have been studied extensively \cite{SCH20} as a natural way of modeling `flows'. In this case, $\mbox{im}(B_k^{\top})$ corresponds to flows induced by gradients on the nodes, $\mbox{ker}(L_k)$ corresponds to curl-free and gradient-free flows, and $\mbox{im}(B_{k+1})$ corresponds to flows that curl around triangles. Such flows on simplicial complexes have been used to model traffic flows \cite{JIA19}.


\section{Simplicial Communities and Spectral Properties of the Hodge Laplacians}
The spectrum of the graph Laplacian encodes important properties about the structure, geometry and dynamics of a network. Indeed it is known that the graph Laplacian encodes for:
\begin{itemize}
	\item \textit{The number of communities} captured by the degeneracy of the zero eigenvalue of the graph Laplacian $L_{[0]}$.
	\item \textit{The identity of communities} captured by the signed support of  the non-zero eigenvectors of the Laplacians when the graph is not symmetric (one can include random weights to the edges to remove the degeneracy of the eigenvalues with probability measure one).
	\item \textit{Community structure} captured by the sign of the eigenvectors associated with small eigenvalues \cite{capocci2005detecting,von2007tutorial}.
	\end{itemize}

While several works have studied community detection in graphs, little attention has been paid to extending this to simplicial complexes. Here we claim that the Hodge Laplacian can generalize the properties of the graph Laplacian as it encodes for the following:

\begin{enumerate}
	\item \textit{The number of  $k$-dimensional cavities or Betti number $\beta_k$} captured by the degeneracy of the zero eigenvalue of the Hodge Laplacian $L_k$.
	\item \textit{The identity of $k$-simplicial communities } is captured by the support of the non-zero eigenvectors of the up-Laplacian $L_k^u$. For symmetric graphs with degenerate eigenvalues, the corresponding eigenvectors become non-unique. In this case, one can include random weights to the edges to remove the degeneracy of the eigenvalues with probability measure one. Since $L_k^{up}$ is isomorphic to $L_{k+1}^{down}$, these are also captured by the support of the non-zero eigenvectors of $L_k^u$. Interestingly if the simplicial complex is the clique complex of a network the $k$-simplicial communities reduce to the $(k+1)$-clique communities of the network. Therefore here we  point out the relation between the clique communities of a network and the spectral properties of the Hodge Laplacian of its clique complex.
\end{enumerate}

While the first property is among the most celebrated of the higher-order Laplacians, the second  has not been sufficiently studied. This work, to the best of our knowledge, is the first investigation of the second property.

\section{Spectral Algorithm for Detection of  Simplicial Communities}

\subsection{The Fundamental Observation} 

The Laplacian in fluid mechanics indicates flux of the gradient of the flow. Hence, the Laplace operator has a physical interpretation as a measure of diffusion. A direct analog of this interpretation exists in simplicial complexes. In particular the  $k$-up Hodge Laplacian $L_k^u$ is encoding diffusion from $k$-simplices to other $k$-simplices if they are $(k+1)$-connected and the $k$-down Hodge Laplacian $L_k^d$ is encoding diffusion from $k$-simplices to other $k$-simplices if they are $(k-1)$-connected. 
From the expression of the off-diagonal matrix elements of $L_k^{u}$ and $L_k^{d}$ (given by Eq. (\ref{Lupe}) and Eq. (\ref{Ldowne}) respectively) it is also apparent that these matrix elements are only non-zero among pairs of $k$-simplices that are upper or lower adjacent, i.e. they are either both faces of the same $(k+1)$-simplex or their intersection is a (non-empty) $(k-1)$-simplex.
 It follows from diffusion laws that there exist a basis of eigenvectors of $L_k^u$ in which the $k$-simplices in the support of each eigenvector belong to a single $k$-simplicial community. Similarly it is also immediate to deduce that  there exist a basis of  eigenvectors of $L_k^d$ in which the $k$-simplices in the support of each eigenvector belong to a single $(k-1)$-simplicial community (or $k$-down simplicial community). 
 Therefore by looking at the spectral properties of the higher-order Hodge Laplacians $L_k^u$ and $L_{k+1}^d$, and in particular by considering the support of their eigenvectors, it is possible to extract the $k$-simplicial communities of a simplicial complex,  that for a clique complex of the network, reduce to  the $(k+2)$-clique communities  of the network \cite{PAL05}.

\begin{figure}[ht]
	\includegraphics[width=0.9\linewidth]{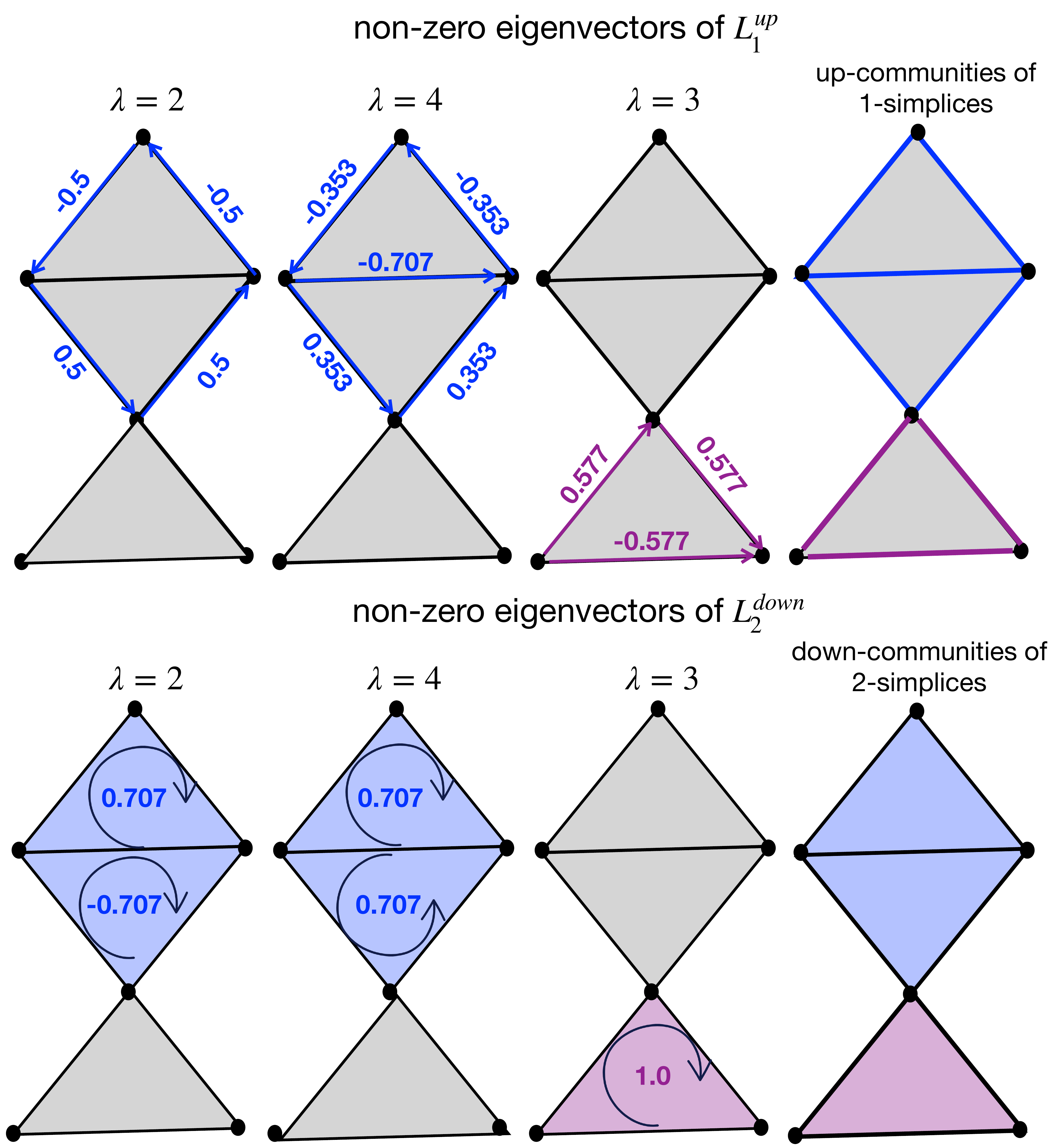}\\
	\caption{Illustrative example of the support of the nonzero eigenvectors $L_1^u$ and $L_2^d$ for a given simplicial complex. (top) Three nonzero eigenvectors of $L_1^{u}$ (encoding diffusion between edges that are upper-adjacent) with localized support on $1$-simplicial communities.The values of the eigenvector components are listed (color coded) next to the simplex (edge) they correspond to. Arrows indicate orientation. (bottom) Nonzero eigenvectors of $L_2^{d}$ (encoding diffusion amongst triangles that are lower adjacent) with localized support on the triangles of $1$-simplicial communities.  Circular arrows indicate orientation (clockwise or anticlockwise). Eigenvector components are listed within the simplex (triangle) they correspond to. Simplices (with localized support) are color coded by their community.  $\lambda$ indicates corresponding eigenvalues.}
	\label{eig}
\end{figure}

\subsection{Algorithm for Identifying $k$-Simplicial Communities}
The $k$-up-Laplacian and the $k$-down Laplacian typically have a highly degenerate zero eigenvalue due to Hodge decomposition in Eq. (\ref{imker}). Therefore although theoretically there is a guarantee that $\mbox{ker}(L_k^u)$ admits a basis formed by vectors each with support in a single $k$ simplicial community, numerically finding this decomposition from the spectrum of $L_k^u$ might be non trivial. 
In this work we formulate an algorithm to best extract the $k$-simplicial communities from the support of the nonzero  eigenvectors of $L_k^u$. 
Labeling the nonzero eigenvalues of $L_k^u$ as
$$\lambda^u_1 \leq \lambda^u_2 \leq \ldots \leq \lambda^u_{\tilde{n}_k},$$ 
let their corresponding eigenvectors be 
$$v^u_1 , v^u_2 , \ldots , v^u_{\tilde{n}_k}.$$
If the eigenvalues are nondegenerate (convert the $\leq$ in the above ordering to $<$), then the support of each eigenvector $v^u$ is localized to the $k$-simplices belonging to a single $k$ simplicial community, i.e.
\bea
sup (v^u_i)  \in \pi_l 
\eea
with 
\bea
sup (v_i^u)\cap \pi_k=\emptyset \ \ \textup{if}\  k\neq l
\eea
where $sup(.)$ denotes the $k$-simplices that form the support the eigenvector. The above equation indicates that the support of a nonzero eigenvector are a subset of only one simplicial community (say $\pi_l$) without being in any other simplicial community $\pi_k$.
Similarly, since $L_k^u$ is isomorphic to $L_{k+1}^d$, the $(k-1)$ simplicial communities are simply the faces of corresponding $k$ simplicial communities.
For example, for $k=1$, one can find  down-communities/clique communities of triangles connected through edges by considering the support of the nonzero eigenvectors of $L_1^u$ instead of $L_2^d$. Fig.~\ref{eig} represent the nonzero eigenvectors of  $L_1^u$ and $L_2^d$ of  a simplicial complex and demonstrates that their support is localized on isomorphic simplicial communities.

Finally we note that the non-zero eigenvectors of the Hodge Laplacian $L_k$ are either nonzero eigenvectors of $L_k^u$ (localized on $k$-simplices belonging to the $k$-up  community) or nonzero eigenvectors of $L_k^d$  (localized on $k$-simplices belonging to the $k$-down community). The eigenvectors corresponding to the zero eigenvalue of $L_k$ have a basis where they are  localized on $k$-dimensional cavities.

On the basis of the above considerations, we  have  formulated the following algorithm to detect the $k$- simplicial communities: 
\begin{enumerate}
	\item Given a graph, compute the boundary matrices for each dimension and store as a sparse matrix.
	\item Compute the corresponding up and down Laplacian through Eq.~(\ref{Lup})
	\item Perform an eigenvector decomposition of $L^u$, and identify the nonzero eigenvectors. 
	\item Compute the support (simplices on which they are localized) of each nonzero eigenvector. If two eigenvectors have overlapping support, take the union of their supports. \textit{Non-overlapping supports indicate different $k$-simplicial communities in absence of network symmetries}.
	\item Visualize $k$-simplicial communities (using networkx and plotly in python)
\end{enumerate}

Given a graph, python code for computing simplicial communities for arbitrary dimensional simplicial complexes is provided at github/chimeraki/Simplicial\_communities. The pseudocode is given in Algorithm \ref{alg:communities}. The computational complexity is constrained by the eigenvector decomposition, which in python LApack has computational complexity of $O(n^3)$. 

One caveat of this approach, as with any spectra-based community detection approach, is that it can be limited by the symmetries of the networks that typically lead to degeneracies of the nonzero eigenvectors. To tackle this possible problem one can devise algorithms to  rotate the corresponding eigenvectors with the goal of  separating the support of the clique communities. In theory, such a rotation always exists. In practice most real world graphs have low symmetry, i.e., if the number of independent eigenvalues are larger than the number of $k$-connected communities. In graphs without \textit{global} symmetry, it is possible to find a basis that reveals the communities. Hence, in general, identification of $k$-communities works well for real graphs based on the non-degenerate eigenvectors alone. Note that simplices of dimension less than $k$ can be the faces of more than one $k$-simplicial communities.

\begin{algorithm}[H]
	\caption{Simplicial community detection via the spectrum of up-Laplacians}
\label{alg:communities}
	\begin{algorithmic} 
		\REQUIRE  $d$-dimensional simplicial complex
		\ENSURE $k$-simplicial communities 
	    \STATE $commList = [\ ]$
		\FOR {$k \text{ from } 1\rightarrow K $}
		\STATE \hspace{0.5em} commList.append([ ])
		\STATE \hspace{0.5em} compute sparse boundary matrices $B_k$
		\STATE \hspace{0.5em} compute $L_k^u$   (or  $L_{k+1}^d$) from Eq. (\ref{Lup})
		\STATE \hspace{0.5em} Find nonzero eigenvectors {$\{v_i^u\} = [v_1^u, v_2^u, \ldots v_{n_{k}}^u]$} of
	 $L_k^u$ (or nonzero eigenvectors $\{v^{d}\} = [v_1^d, v_2^d, \ldots v_{n_{k+1}}^d]$ of $L_{k+1}^d$)
	 	\STATE \hspace{0.5em} Initialize $w_i^{(0)}=sup(v_i^u)$ (or initialize $w_i^{(0)}=sup(v_i^d)$)
		\STATE \hspace{0.5em} \textbf{if} $w_i^{(r)}\cap w_j^{(r)} \neq \nullity$ and $\lambda_i = \lambda_j$ \textbf{then}
				\STATE 		\hspace{1em} 	Take union of overlapping supports\
				\STATE \hspace{1em}  $w_i^{(r+1)} = w_j^{(r+1)} = w_i^{(r)} \cup w_j^{(r)}$
				\STATE  \hspace{0.5em} commList[k].append({$w$})
		\STATE \hspace{0.5em} Visualize communities
		\ENDFOR
	\end{algorithmic}
\end{algorithm}

\section{Higher-order Simplicial Communities}

We present the results on $k$-simplicial communities identified through decomposing the support of the nonzero eigenvectors of the $k^{th}$ up-Laplacian. As a consequence of the isomorphism between $L_k^u$ and  $L_{k+1}^d$, the $(k+1)$-down communities are isomorphic to the $k$-up communities, i.e., the faces of the $(k+1)$-down communities form up-communities of $k$-simplices. The simplices of each down simplicial community is color coded by community. We present results for $k=1$ and $k=2$ for several simplicial complexes with varying levels of symmetry, and with and without holes. Interestingly, for cases with holes, the harmonic representatives themselves are in the kernel of $L_k^u$, however, we are still able to find simplicial communities. 
\begin{figure}[htbp!]
	\includegraphics[width=\linewidth]{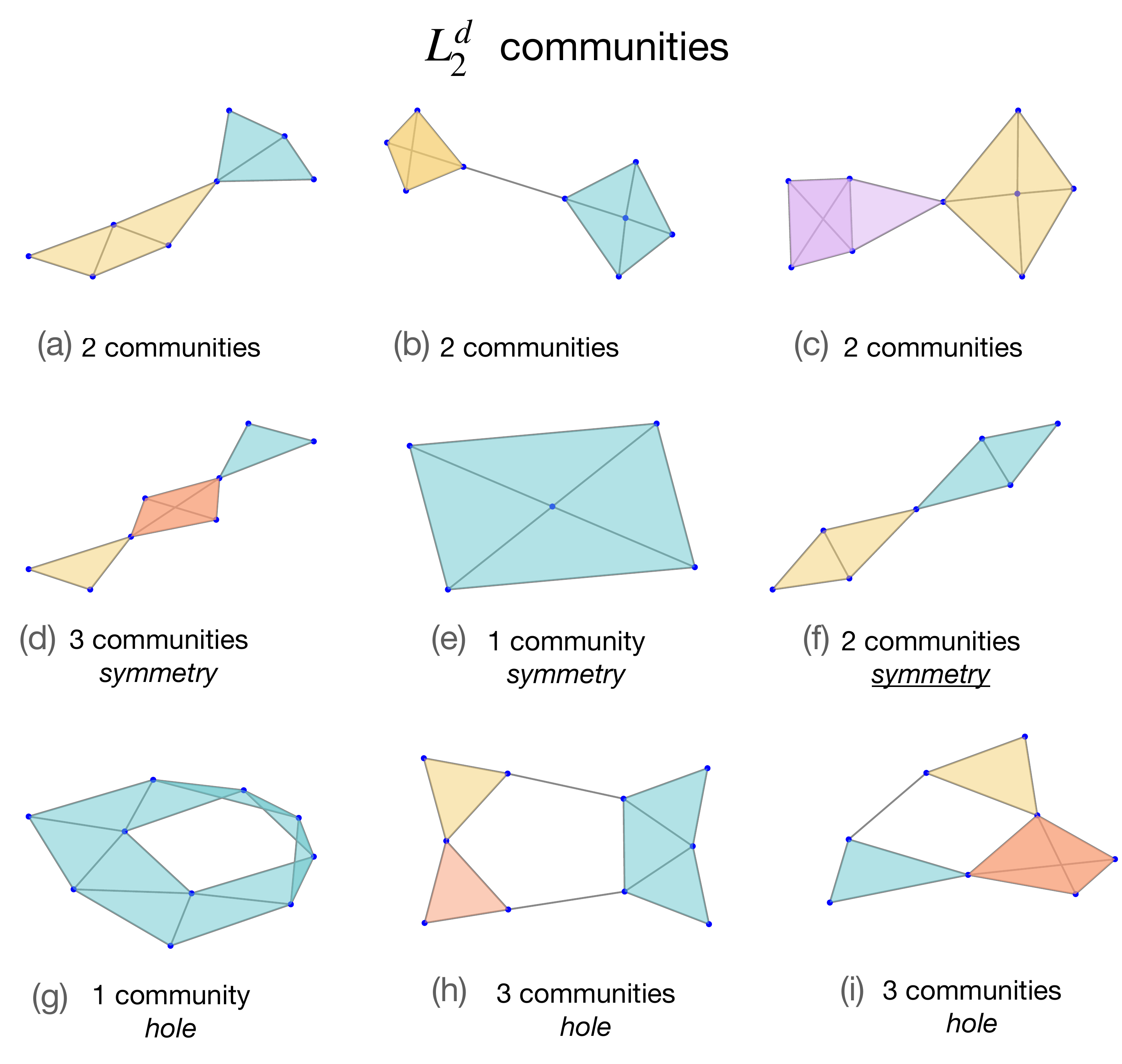}
		\caption{Color-coded 2-down-communities obtained from the spectrum of the 2-down Laplacian. All edges have unit weight.}
	\label{fig:2-communities}
\end{figure}

\subsection{2-Down Communities}
Figures \ref{fig:2-communities}(a)-(c) show simplicial complexes with non-symmetric simplicial communities, Figs. \ref{fig:2-communities}(d)-(f) show simplicial complexes with symmetric simplicial communities and Figs. \ref{fig:2-communities}(g)-(i) show simplicial complexes with non-trivial Betti numbers.

There exist several algorithms that find sparse eigenvectors which typically separate the support when degeneracy arises \cite{GUE21}. In this work, we use LApack  functions built into the sklearn package in python. Grouping of communities is typically observed only when the entire network displays symmetry (as opposed to symmetry in a small subset of simplices). In a large majority of networks, global symmetries are rare.

\subsection{3-Down Communities}

\begin{figure}[htbp!]
	\includegraphics[width=\linewidth]{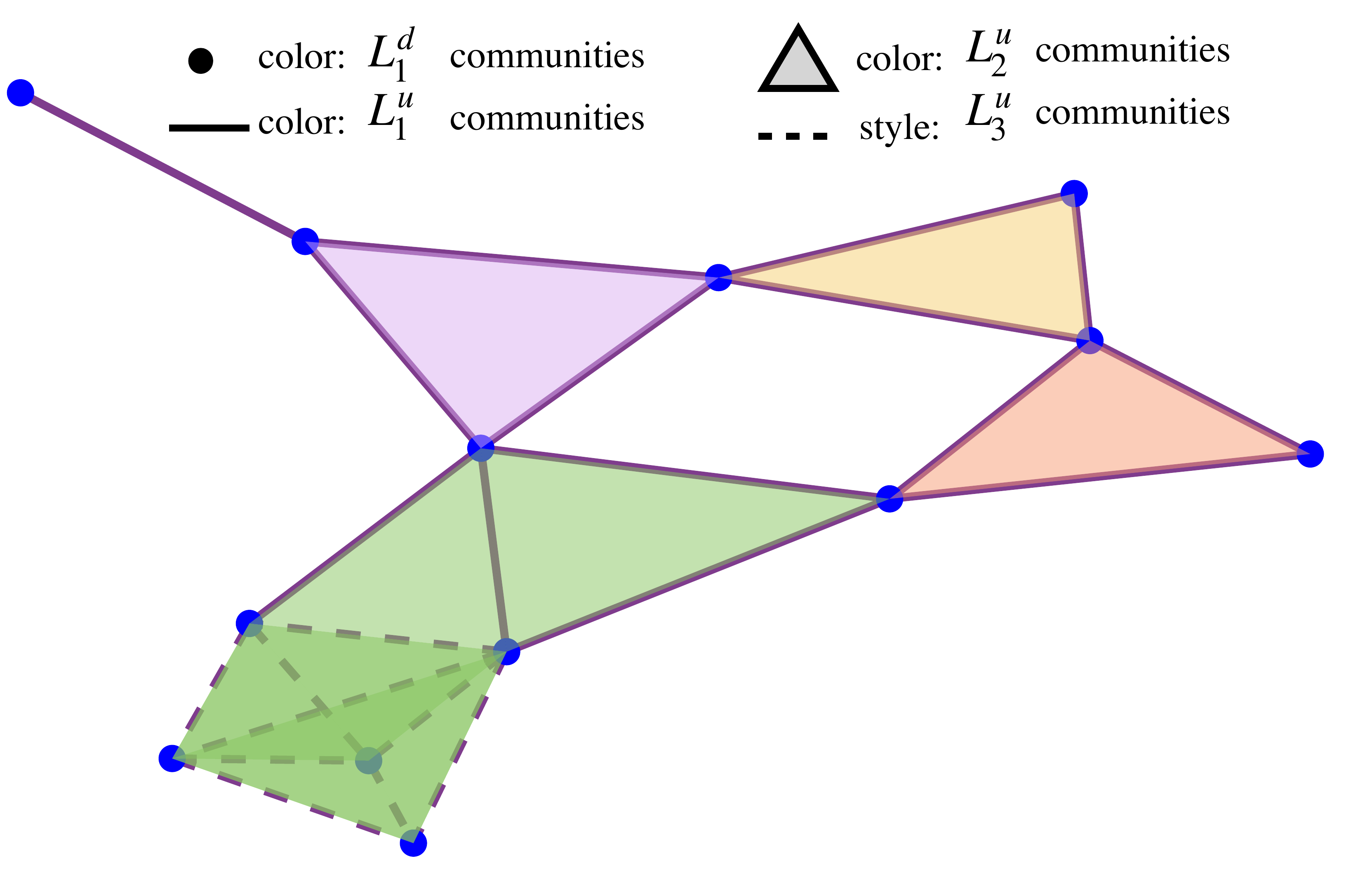}
		\caption{Color-coded communities for 3 different dimensions. 1-clique communities obtained from the spectrum of the 0-up Laplacian color coded on the 1-simplices (edges). Four 2-clique communities obtained from the spectrum of the 1-up Laplacian color coded on the 2-simplices (triangles). One 3-clique community obtained from the spectrum of the 2-up Laplacian marked through dashed edges. All edges have unit weight.}
	\label{fig:3-communities}
\end{figure}

Spectral community detection in simplicial complexes method can be  extended to arbitrary dimension $k \geq 2$. In Fig. \ref{fig:3-communities}, we show the higher-order simplicial communities for $k=0,1,2$. For better visualization of the simplicial communities we limit the dimension of the simplicial complex to three. The 1-down communities (edges connected by nodes) are indicated through edge coloring. There exists only one 1-down community since the network skeleton is fully connected. The 1-down communities also encode the 0-up communities (nodes connected by edges - also known as connected components of the network). 
\\

The 2-down (or 1-up) communities (triangles connected by edges are marked through coloring on the faces of triangles). There are four such simplicial communities in number. Lastly,The 3-down communities (tetrahedron connected by overlapping triangle faces) are marked through all the edge faces of tetrahedron in a community denoted by a specific style. In Fig \ref{fig:3-communities}, there exists one 3-down community (obtained from the support of nonzero eigenvectors of $L_3^u$) comprising of 2 tetrahedrons denoted by dashed edges.

\begin{figure*}[htp!]
	\includegraphics[width=2.0\columnwidth]{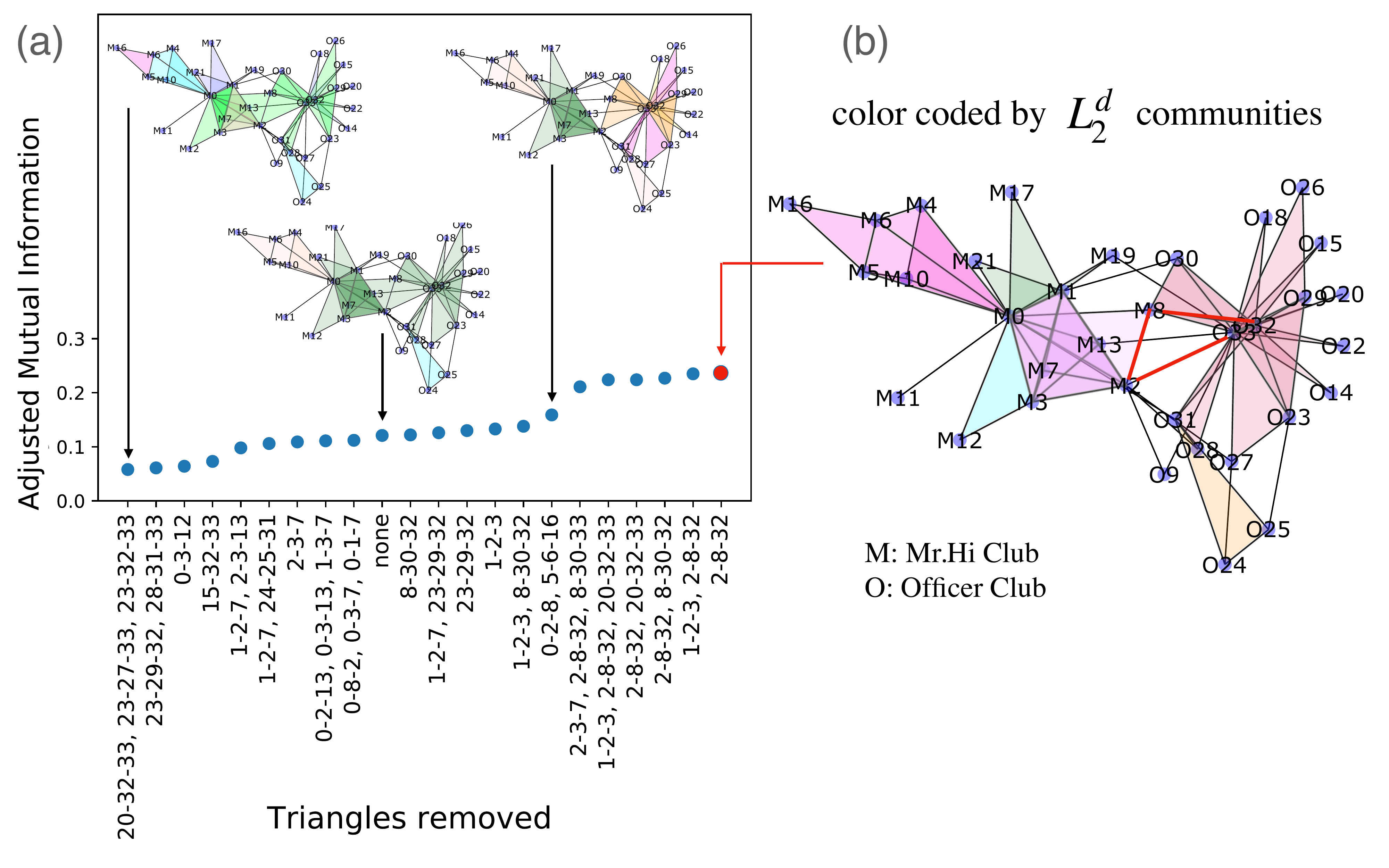}
	\caption{(a) AMI plotted across several cases of unfilled 2-simplices (each triangle is labelled by a collection of $3$-nodes on the $x$-axis). The highest AMI was obtained upon removing the simplex $2$-$8$-$32$. AMI is averaged over $100$ samples for each configuration. The standard error is of the order $0.0001$. (b) The configuration with the highest AMI is shown here. The removed simplex (unfilled triangle) is marked with red edges. The $2$-down simplicial  communities of the Zachary Karate Club network are color-coded. The first part of the label indicate the club affiliation (clubs are indicated by M and O), and the second part is a numerical identifier of the individual. Hence simplex removed ($2$-$8$-$32$) was a $3$-way connection between player $2$ in club M, player $8$ in club M, and player $32$ in Club O. Note that the labels of the nodes in the $x$-axis are the numerical identifier of each node, 
e.g. M16 is uniquely identified by its numerical part $16$. Node positions are determined through the Kamada Kawai layout for weighted graphs.}
	\label{fig:karate}
\end{figure*}

\section{Simplicial Communities of Real Networks}

While several works have considered community detection based on spectral clustering of the graph Laplacian $L_{[0]}$, many real world networks such as neuronal networks, social interaction networks, transportation networks etc. study flows along edges. higher-order relationships, best encoded through simplicial complexes, provide crucial insight into connectivity of the simplicial complex and are becoming increasingly important for analyzing real networks. While there is a dearth of publicly available datasets for simplicial complexes, one can generate a simplicial from a network skeleton considering the clique complex or alternatively considering other simplicial complexes admitting the original network as their skeleton. Here we present simplicial community detection based on $L_1^u$ (or $L_2^d$) in order to obtain simplicial communities of  edges that are triangle connected or equivalently of triangles connected by edges. Such communities are indicative of localization of edge flows within communities, providing important information about the nature of signal propagation in the simplicial complex.

\subsection{Extracting True Simplices: Karate Club Network}

We present an example of social interactions in a Karate Club network.
The Zachary Karate Club is a network of interactions amongst $34$ members of a karate club outside the club between $1970$ to $1972$. The original club eventually split up into two clubs: Officer denoted as `O' and Mr. Hi denoted as `M' (pseudonyms).
The dataset became a popular example of community detection after being used in \cite{GIR02}.

It is important to note that datasets for simplicial complexes are rare. Instead, simplicial complexes are commonly created from the network skeleton. Datasets such as the Karate Club Network do not implicitly come with information about many-body interactions (triangles, tetrahedra etc.). Naively, one may assume each higher-order structure to be filled (e.g. a filled triangle denotes a $2$-simplex), however this may not be an accurate representation of th true higher-order connections. Here we use the knowledge about the true community structure of the network for proposing an algorithm that   extract the true many body interactions. In particular, in order to detect  the triangles that are filled (the true $3$-way interaction as opposed to $3$ different $2$-way interactions), we use the adjusted mutual information \cite{MEI98}. 

In probability theory and information theory, the mutual information (MI) \cite{SHA01} of two random variables is a measure of the mutual dependence between the two variables. Adjusted Mutual Information (AMI) is used to compare how similar two clusters or partitions of data are, adjusting for the effect of agreement due to chance.
The AMI between two partitions $(C, \tilde{C}) $is given by: 
\begin{equation}
	AMI(C, \tilde{C}) = \frac{MI (C, \tilde{C}) - \mathbb{E}[MI(C, \tilde{C})]}{max( H(C)H(\tilde{C})) - \mathbb{E}[MI(C, \tilde{C})]}
\end{equation}
where $H(C)$  is the entropy of partition $C$ given by 
\begin{equation}
	H(C)= -\sum_i^N P_C(i) \log P_C(i)
	\end{equation}
where $N$ is the total number of clusters in $C$ and $P_C(i)$ the probability that a random object from the set falls in cluster $i$ of $C$. AMI takes a value of $1$ when the two partitions are identical and $0$ when the MI between two partitions equals the value expected due to chance alone. 

The clubs affiliations (O vs H) of each individual in the Karate club network is known. Let's call this partition of players $C$. We then compare this to the partition of individuals induced through simplicial communities. It is unclear how many of the triangles in the simplicial complex are filled, hence we run experiments with various possibilities of randomly unfilling one or more triangles (removing $2$-simplices), and retain the structure which partitions the nodes to obtain the highest AMI compared with the partition $C$. 
Note, considering $1$-simplicial communities  gives us communities of edges, however we are interested in clustering nodes in order to use AMI. A single individual can then be in multiple communities. Hence, we average over several partitions where in each partition, affiliations for each node is sampled from the set of communities it belongs to. We pick a large number of samples ($100$) resulting in a low variance in the AMI.

In Fig. \ref{fig:karate}(a) we show the AMI (sorted) for several random configurations of removing one, two or three $2$-simplices. The x-axis denotes the triangles removed by their set of 3-nodes (where the numerical value is used to identify the nodes). The highest AMI was obtained by removing a single simplex ($3$-way connection between player $2$ in club M, player $8$ in club M, and player $32$ in Club O). In (b), we analyze the 1-simplicial communities corresponding to this configuration (with the single removed triangle outlined in red). This identifies $3$-way interactions (modeled by triangles) with at least one overlapping pairwise interaction (edge). We observe that the $1$-simplicial communities naturally split into the M club and the O club. Additionally, there exists a deeper level of higher-order structure within team. The O club has a much larger single 1-simplicial community (in yellow) and a smaller one (in red) comprising of just $3$-members, indicating that there was higher $3$-way interaction on average with the exception of one group of $3$ members. The M club on the other hand is comprised of several $1$-simplicial communities of about comparable size.

\subsection{Persistent Communities: Scientific Collaboration Network}
\begin{figure}[htbp!]
	\includegraphics[width=0.9
	\linewidth]{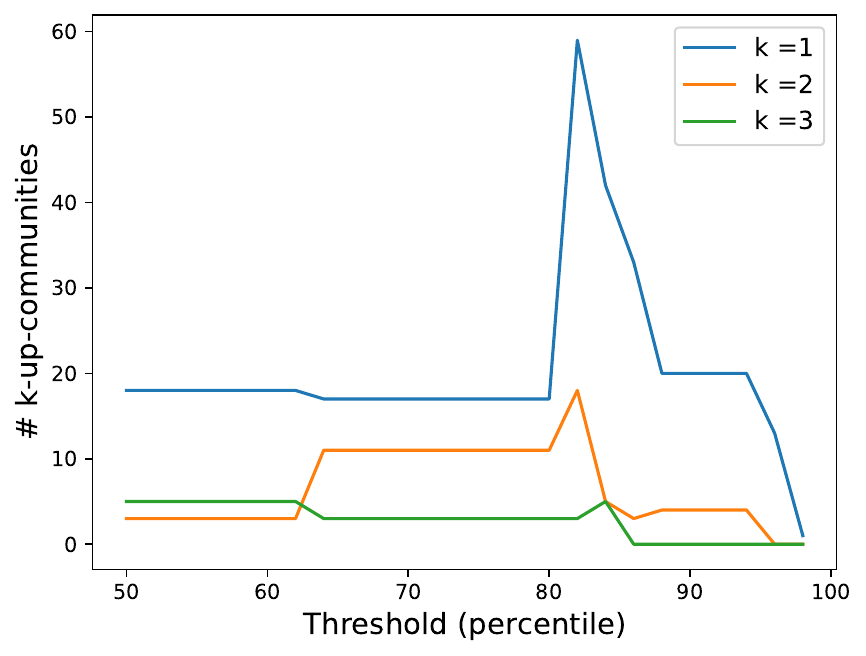}
	\caption{Number of 1-simplicial communities as a function of filtration parameter $\nu$.}
	\label{fig:filtration}
\end{figure}

We also study a collaboration network where nodes are researchers that publish in the field of network science. Here, we introduce how filtrations, that are popular  topological techniques of applied topology, can be applied for the investigation of  simplicial community detection. The data contains a collaboration network of scientists working on network theory and experiment, as compiled in \cite{NEW06}, which also describes the mechanism for setting weights. The network is weighted, and undirected. The original network consists of 1589 vertices and 2742 edges, however as an illustrative example of persistence approaches, we threshold the network and compute simplicial communities across different levels of a `filtration parameter'. Analogous to its definition in persistence topology \cite{petri2014homological}, the filtration parameter is a threshold that is set to a percentile $\nu$ of the edge weight distribution, weights below this are set to zero. Fig.\ref{fig:filtration} shows the number of 1-simplicial communities as a function of the filtration parameter.

We then investigate higher-order community structure at a filtration parameter of 0.8, i.e., only the top 20 percentile of all edge weights are nonzero. This leads to a graph with 84 nodes and 107 edges. We convert this graph to a simplicial complex by filled all triangles, tetrahedra etc. to generate higher-order simplices. The corresponding simplicial complex contains simplicial communities across a wide range of $k$, and are `persistent' communities in the sense that they persist despite high thresholding. Removal of unconnected nodes results in a graph with 114 nodes and 96 edges. 
Similar techniques are used in persistent homology to identify persistence of topological properties at different scales. In particular, \cite{RIE17} investigates clique community persistence, which is closely related to simplicial community persistence.

The 1-up/2-down communities of pairwise collaborations that are connected through 2-simplices (triangles) are listed below:

\begin{enumerate}
	\item Almaas E,  Arenas A, Bennaim E, Burns G, Cabrales A, Diaz-Guilera A, Guimera R, Hilgetag C, Krapivsky P, Newman M, Oneill M, Redner S, Rodgers G, Scannell J, Vegaredondo F, Watts D, Young M 
	\item  Albert R, Barabasi A, Jeong H, Neda Z, Oltvai Z, Ravasz E, Schubert A, Vicsek T 
	\item  Moreno Y, Pastor-Satorras R, Vazquez A, Vespignani A 
	\item Arenas A, Barabasi A, Cabrales A, Danon L, Diaz-Guilera A, Guimera R, Jeong H, Neda Z, Ravasz E, Schubert A, Vega-Redond F, Vicsek T ]
	\item  Albert R, Almaas E, Barabasi A, Bennaim E, Dodds P, Jeong H, Krapivsky P, Moore C, Neda Z, Newman M, Oltvai Z, Ravasz E, Redner S, Rodgers G, Schubert A, Strogatz S, Vicsek T, Watts D 
	\item  Barrat A, Barthelemy M, Moreno Y, Pastor-Satorras R, Vazquez A, Vespignani A 
	\item  Dodds P, Moore C, Newman M, Strogatz S, Watts D 
	\item  Albert R, Almaas E, Amaral L, Arenas A, Barabasi A, Barrat A, Barthelemy M, Bennaim E, Cabrales A, Caldarelli G, Danon L, Diaz-Guilera A, Dodds P, Dunne J, Guimera R, Holme P, Jeong H, Kim B, Krapivsky P, Moore C, Moreno Y, Neda Z, Newman M, Oltvai Z, Pastor-Satorras R, Ravasz E, Redner S, Rodgers G, Schubert A, Stanley H, Strogatz S, Trusina A, Vazquez A, Vega-Redond F, Vespignani A, Vicsek T, Watts D, Williams R 
	\item Arenas A, Burns G, Cabrales A, Diaz-Guilera A, Guimera R, Hilgetag C, Krapivsky P, Oneill M, Redner S, Scannell J, Vega-Redond F, Young M 
	\item  Almaas E, Arenas A, Bennaim E, Cabrales A, Diaz-Guilera A, Dodds P, Guimera R, Krapivsky P, Moore C, Newman M, Redner S, Rodgers G, Strogatz S, Vega-Redondo F, Watts D 
	\item  Barabasi A, Jeong H, Neda Z, Ravasz E, Schubert A, Vicsek T 
	\item  Arenas A, Burns G, Cabrales A, Diaz-Guilera A, Guimera R, Hilgetag C, Oneill M, Scannell J, Vega-Redondo F, Young M 
	\item  Almaas E, Bennaim E, Dodds P, Krapivsky P, Moore C, Newman M, Redner S, Rodgers G, Strogatz S, Watts D 
	\item Barabasi A, Barrat A, Barth\'elemy M, Caldarelli G, Jeong H, Moreno Y, Neda Z, Oltvai Z, Pastor-Satorras R, Ravasz E, Schubert A, Vazquez A, Vespignani A, Vicsek T 
	\item Almaas E, Arenas A, Bennaim E, Burns G, Cabrales A, Diaz-Guilera A, Guimera R, Hilgetag C, Krapivsky P, Oneill M, Redner S, Rodgers G, Scannell J, Vega-Redond F, Young M 
	\item Amaral L, Arenas A, Barabasi A, Barth\'elemy M, Cabrales A, Danon L, Diaz-Guilera A, Dunne J, Guimera R, Jeong H, Neda Z, Oltvai Z, Ravasz E, Schubert A, Stanley H, Vega-Redondo F, Vicsek T, Williams R 
\end{enumerate}

	The 2-up/3-down simplicial  communities of researchers that are faces of triangles (3-way collaboration) and are connected by tetrahedra (4-way collaboration) are listed below. There are 9 communities in total. Note that an individual researcher can be in multiple communities. 
	
\begin{enumerate}
	\item  Arenas A, Barabasi A, Cabrales A, Danon L, Diaz-Guilera A, Guimera R, Jeong H, Neda Z, Oltvai Z, Ravasz E, Schubert A, Vega-Redondo F, Vicsek T 
	\item  Moreno Y, Pastor-Satorras R, Vazquez A, Vespignani A 
	\item  Arenas A, Barab\'asi A, Cabrales A, Danon L, Diaz-Guilera A, Guimera R, Jeong H, Neda Z, Ravasz E, Schubert A, Vega-Redondo F, Vicsek T 
	\item  Barrat A, Barthelemy M, Moreno Y, Pastor-Satorras R, Vazquez A, Vespignani A 
	\item  Arenas A, Burns G, Cabrales A, Diaz-Guilera A, Guimera R, Hilgetag C, Oneill M, Scannell J, Vega-Redondo F, Young M 
	\item  Arenas A, Barabasi A, Burns G, Cabrales A, Danon L, Diaz-Guilera A, Guimera R, Hilgetag C, Jeong H, Neda Z, Oneill M, Ravasz E, Scannell J, Schubert A, Vega-Redondo F, Vicsek T, Young M 
	\item  Arenas A, Cabrales A, Diaz-Guilera A, Guimera R, Vega-Redond F 
	\item  Barabasi A, Jeong H, Neda Z, Oltvai Z, Ravasz E, Schubert A, Vicsek T 
	\item  Arenas A, Cabrales A, Danon L, Diaz-Guilera A, Guimera R, Vega-Redond F 
\end{enumerate}

	The 3-up/4-down simplicial  communities of researchers that are faces of tetrahedra (4-way collaboraion) that are connected through 4-simplices are listed below. There are 9=3 communities in total. Note that an individual researcher can be in multiple communities. 
	
	\begin{enumerate}
		\item  Arenas A, Cabrales A, Diaz-Guilera A, Guimera R, Vega-Redond F 
		\item  Burns G, Hilgetag C, Oneill M, Scannell J, Young M 
		\item  Barabasi A, Jeong H, Neda Z, Ravasz E, Schubert A, Vicsek T 
		\end{enumerate}
	
\begin{figure*}[htbp!]
	\includegraphics[width=1
	\linewidth]{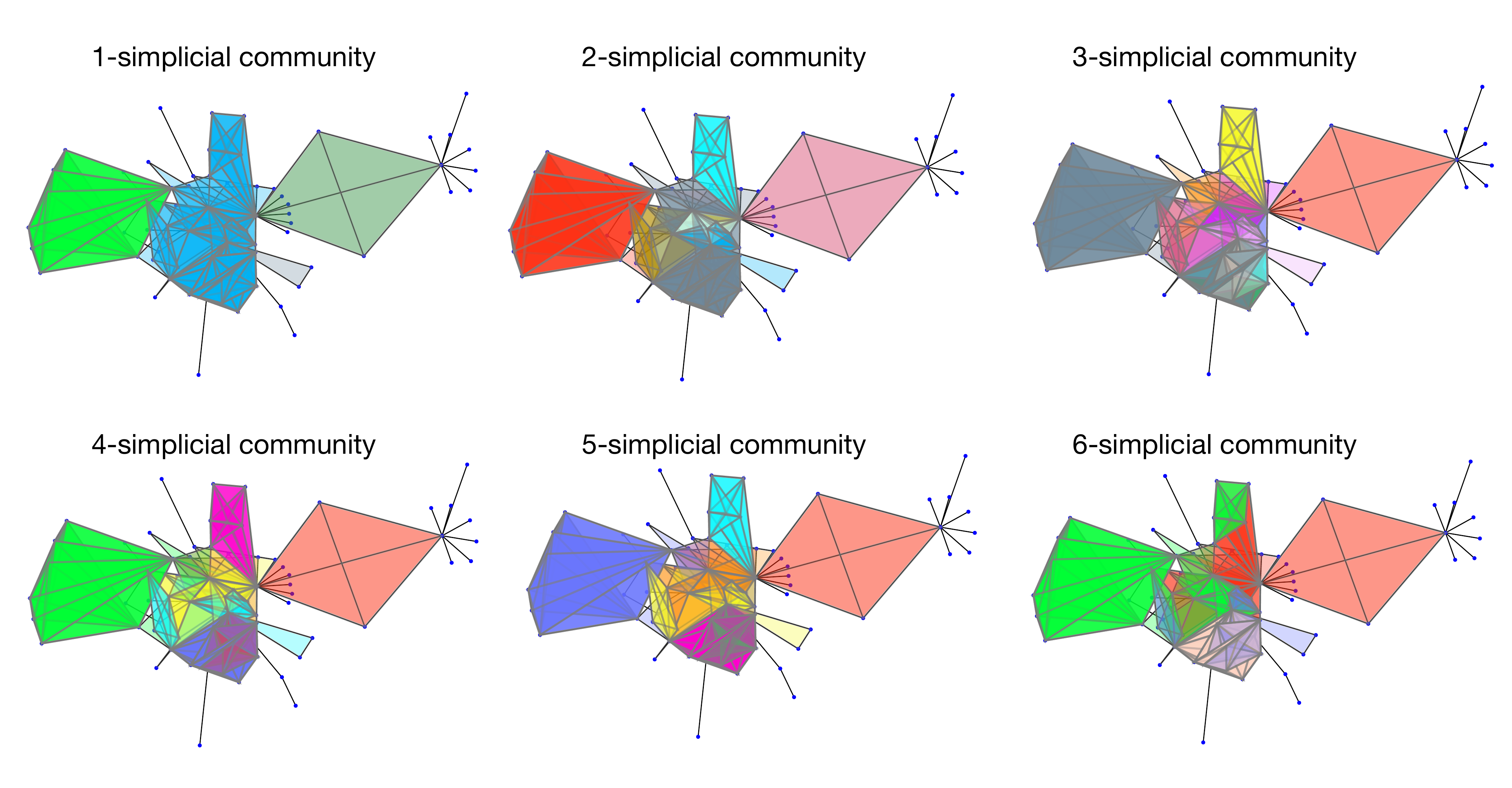}
	\caption{Illustrative visualization of the $k$-simplicial communities color-coded by their triangles of the Les Miserable word network. Figure is illustrative and nodes aren't listed (see Appendix \ref{app:lesmiscomm} for list of character affiliations). Plots are presented for various dimensional simplicial communities from $k=1$ upto $k=6$.}
	\label{fig:les_mis}
\end{figure*}

\subsection{Higher Dimensional Communities in Language: Les Mis\'erables Network}

The use of network approaches for analysis of word association networks and for natural language processing has gained tremendous impetus in recent years \cite{VIT15,WIL02}. In fact, language and literally naturally contain layered structure, making them suitable candidates for simplicial analysis. For instance, characters in a book often tend to have nuanced higher-order interactions at different scales, corresponding to the existence of higher-order simplices.

Fig. \ref{fig:les_mis} presents a visualization of the simplicial communities of simplicial complexes obtained from the word association network that encodes relationships between characters in Victor Hugo’s novel, Les Mis\'erables. It contains $77$ vertices corresponding to characters of the novel, and $254$ edges connecting two characters whenever they appear in the same chapter. Edge weight between two words indicates the number of times they co-appear in the same sentence. The network is found to contain simplices of dimension up to $k=6$. The simplicial communities for varying $k$ are visualized through their projection on 2-simplices/triangles. For instance, 1-simplicial communities (edges connected by triangles) are captured through the color of the triangle. 2-simplicial communities (triangles connected by tetrahedra) are color coded in the triangles, 3-simplicial communities (tetrahedra connected by 4-simplices) are identifiable through the color-coding of the triangle-faces of all tetrahedra within a particular community etc. The figure is for illustrative purposes lacks node-labelling for convenient visualization, however a detailed list of the individuals belonging to the simplicial community for all $k=1,2,3,4,5,6$ are provided in Appendix \ref{app:lesmiscomm}. A reader of the book may notice expected patterns and community structure in higher-order interaction in the list.

\section{Conclusions}

Higher-order interactions are ubiquitous and are increasingly recognized as an an important feature of complex systems, yet are largely neglected. Higher-order interactions can be captured by simplicial complexes that are the  building blocks of discrete topologies. Therefore, describing a complex system as a simplicial complex allows for its analysis through the powerful tools of algebraic topology, enabling investigation of its topological invariants such as Betti numbers. This line of research has given rise to a prosperous and growing field at the interface between Topological Data Analysis and Network Science. However, from a Network Science perspective, several unanswered questions could benefit from the deeper insight and unique perspectives obtained by a simplicial complex representation of higher-order networks.

A fundamental question is how a higher-order network can be partitioned into communities. Here we propose to partition the higher-order simplicies of a simplicial complex in $k$-simplicial communities, where all simplices within a community are adjacent. 
The relation between $k$-simplicial communities and the spectral properties of the Hodge Laplacian of the simplicial complex is exploited to propose a spectral algorithm for simplicial community detection. The simplicial communities are identified by the support of the eigenvectors which intuitively encodes diffusion amongst simplices through higher/lower dimensional simplicial faces. Through Hodge decomposition, we interpret these in terms of higher-order curls and gradients, providing an explanation of flow patterns in simplicial complexes.

When the simplicial complex under consideration is the clique complex of a network, the notion of simplicial communities can be related to the notion of clique communities.
However, in this work we highlight the difference between the simplicial community and the clique community of the skeleton of the simplicial complex. The clique complex of the network skeleton is {\em not} in general the equal to the original simplicial complex. In fact the clique complex of the network skeleton of a simplicial complex, in general, include at least as many, and typically more, higher dimensional simplicies than the original simplicial complex.
Having noted this difference we use simplicial communities to infer the true higher-order interactions from the knowledge of a network and its true community structure. Simplicial data is very scarce, and such techniques are increasingly being investigated to convert network data into simplicial complex. In particular we apply this inference algorithm to infer the true $3$-way interactions in the famous Zachary Karate Club.

Our study of simplicial communities of real networks is also extended to weighted collaboration networks where show that simplicial communities can be studied as a function of the opportune filtration of the simplicial complex performed by thresholding all the links with a tunable threshold.
This allow us to define persistent communities, i.e. set of $k$-simplices remaining $(k-1)$-connected for a large interval of values of the threshold.

Finally we also argue that simplicial communities are an important features of the increasing number of network datasets extracted by natural language processing algorithms. As an example we provide the case of the simplicial communities of the characters of the book Les Mis\'erables.

In conclusion our work show that simplicial communities are fundamental structural features of simplicial complexes that are encoded in their higher-order spectral properties. The study of simplicial communities of real networks can be used to infer true higher-order interactions and persistent communities, i.e. communities that remains unchanged upon different thresholding of the weight of the network.

\bibliography{simplicial_comm}

\begin{thebibliography}{74}%
\makeatletter
\providecommand \@ifxundefined [1]{%
 \@ifx{#1\undefined}
}%
\providecommand \@ifnum [1]{%
 \ifnum #1\expandafter \@firstoftwo
 \else \expandafter \@secondoftwo
 \fi
}%
\providecommand \@ifx [1]{%
 \ifx #1\expandafter \@firstoftwo
 \else \expandafter \@secondoftwo
 \fi
}%
\providecommand \natexlab [1]{#1}%
\providecommand \enquote  [1]{``#1''}%
\providecommand \bibnamefont  [1]{#1}%
\providecommand \bibfnamefont [1]{#1}%
\providecommand \citenamefont [1]{#1}%
\providecommand \href@noop [0]{\@secondoftwo}%
\providecommand \href [0]{\begingroup \@sanitize@url \@href}%
\providecommand \@href[1]{\@@startlink{#1}\@@href}%
\providecommand \@@href[1]{\endgroup#1\@@endlink}%
\providecommand \@sanitize@url [0]{\catcode `\\12\catcode `\$12\catcode
  `\&12\catcode `\#12\catcode `\^12\catcode `\_12\catcode `\%12\relax}%
\providecommand \@@startlink[1]{}%
\providecommand \@@endlink[0]{}%
\providecommand \url  [0]{\begingroup\@sanitize@url \@url }%
\providecommand \@url [1]{\endgroup\@href {#1}{\urlprefix }}%
\providecommand \urlprefix  [0]{URL }%
\providecommand \Eprint [0]{\href }%
\providecommand \doibase [0]{http://dx.doi.org/}%
\providecommand \selectlanguage [0]{\@gobble}%
\providecommand \bibinfo  [0]{\@secondoftwo}%
\providecommand \bibfield  [0]{\@secondoftwo}%
\providecommand \translation [1]{[#1]}%
\providecommand \BibitemOpen [0]{}%
\providecommand \bibitemStop [0]{}%
\providecommand \bibitemNoStop [0]{.\EOS\space}%
\providecommand \EOS [0]{\spacefactor3000\relax}%
\providecommand \BibitemShut  [1]{\csname bibitem#1\endcsname}%
\let\auto@bib@innerbib\@empty
\bibitem [{\citenamefont {Barab{\'a}si}(2016)}]{BAR13}%
  \BibitemOpen
  \bibfield  {author} {\bibinfo {author} {\bibfnamefont {A.-L.}\ \bibnamefont
  {Barab{\'a}si}},\ }\href@noop {} {\emph {\bibinfo {title} {Network
  science}}}\ (\bibinfo  {publisher} {Cambridge University Press},\ \bibinfo
  {year} {2016})\BibitemShut {NoStop}%
\bibitem [{\citenamefont {Telesford}\ \emph {et~al.}(2011)\citenamefont
  {Telesford}, \citenamefont {Simpson}, \citenamefont {Burdette}, \citenamefont
  {Hayasaka},\ and\ \citenamefont {Laurienti}}]{TEL11}%
  \BibitemOpen
  \bibfield  {author} {\bibinfo {author} {\bibfnamefont {Q.~K.}\ \bibnamefont
  {Telesford}}, \bibinfo {author} {\bibfnamefont {S.~L.}\ \bibnamefont
  {Simpson}}, \bibinfo {author} {\bibfnamefont {J.~H.}\ \bibnamefont
  {Burdette}}, \bibinfo {author} {\bibfnamefont {S.}~\bibnamefont {Hayasaka}},
  \ and\ \bibinfo {author} {\bibfnamefont {P.~J.}\ \bibnamefont {Laurienti}},\
  }\href@noop {} {\bibfield  {journal} {\bibinfo  {journal} {Brain
  connectivity}\ }\textbf {\bibinfo {volume} {1}},\ \bibinfo {pages} {295}
  (\bibinfo {year} {2011})}\BibitemShut {NoStop}%
\bibitem [{\citenamefont {Hogan}\ \emph {et~al.}(2008)\citenamefont {Hogan},
  \citenamefont {Fielding}, \citenamefont {Lee} \emph {et~al.}}]{HOG08}%
  \BibitemOpen
  \bibfield  {author} {\bibinfo {author} {\bibfnamefont {B.}~\bibnamefont
  {Hogan}}, \bibinfo {author} {\bibfnamefont {N.}~\bibnamefont {Fielding}},
  \bibinfo {author} {\bibfnamefont {R.}~\bibnamefont {Lee}},  \emph {et~al.},\
  }\href@noop {} {\bibfield  {journal} {\bibinfo  {journal} {The Sage handbook
  of online research methods}\ ,\ \bibinfo {pages} {141}} (\bibinfo {year}
  {2008})}\BibitemShut {NoStop}%
\bibitem [{\citenamefont {Gosak}\ \emph {et~al.}(2018)\citenamefont {Gosak},
  \citenamefont {Markovi{\v{c}}}, \citenamefont {Dolen{\v{s}}ek}, \citenamefont
  {Rupnik}, \citenamefont {Marhl}, \citenamefont {Sto{\v{z}}er},\ and\
  \citenamefont {Perc}}]{GOS18}%
  \BibitemOpen
  \bibfield  {author} {\bibinfo {author} {\bibfnamefont {M.}~\bibnamefont
  {Gosak}}, \bibinfo {author} {\bibfnamefont {R.}~\bibnamefont
  {Markovi{\v{c}}}}, \bibinfo {author} {\bibfnamefont {J.}~\bibnamefont
  {Dolen{\v{s}}ek}}, \bibinfo {author} {\bibfnamefont {M.~S.}\ \bibnamefont
  {Rupnik}}, \bibinfo {author} {\bibfnamefont {M.}~\bibnamefont {Marhl}},
  \bibinfo {author} {\bibfnamefont {A.}~\bibnamefont {Sto{\v{z}}er}}, \ and\
  \bibinfo {author} {\bibfnamefont {M.}~\bibnamefont {Perc}},\ }\href@noop {}
  {\bibfield  {journal} {\bibinfo  {journal} {Physics of life reviews}\
  }\textbf {\bibinfo {volume} {24}},\ \bibinfo {pages} {118} (\bibinfo {year}
  {2018})}\BibitemShut {NoStop}%
\bibitem [{\citenamefont {Havlin}\ \emph {et~al.}(2012)\citenamefont {Havlin},
  \citenamefont {Kenett}, \citenamefont {Ben-Jacob}, \citenamefont {Bunde},
  \citenamefont {Cohen}, \citenamefont {Hermann}, \citenamefont {Kantelhardt},
  \citenamefont {Kert{\'e}sz}, \citenamefont {Kirkpatrick}, \citenamefont
  {Kurths} \emph {et~al.}}]{HAV12}%
  \BibitemOpen
  \bibfield  {author} {\bibinfo {author} {\bibfnamefont {S.}~\bibnamefont
  {Havlin}}, \bibinfo {author} {\bibfnamefont {D.~Y.}\ \bibnamefont {Kenett}},
  \bibinfo {author} {\bibfnamefont {E.}~\bibnamefont {Ben-Jacob}}, \bibinfo
  {author} {\bibfnamefont {A.}~\bibnamefont {Bunde}}, \bibinfo {author}
  {\bibfnamefont {R.}~\bibnamefont {Cohen}}, \bibinfo {author} {\bibfnamefont
  {H.}~\bibnamefont {Hermann}}, \bibinfo {author} {\bibfnamefont
  {J.}~\bibnamefont {Kantelhardt}}, \bibinfo {author} {\bibfnamefont
  {J.}~\bibnamefont {Kert{\'e}sz}}, \bibinfo {author} {\bibfnamefont
  {S.}~\bibnamefont {Kirkpatrick}}, \bibinfo {author} {\bibfnamefont
  {J.}~\bibnamefont {Kurths}},  \emph {et~al.},\ }\href@noop {} {\bibfield
  {journal} {\bibinfo  {journal} {The European Physical Journal Special
  Topics}\ }\textbf {\bibinfo {volume} {214}},\ \bibinfo {pages} {273}
  (\bibinfo {year} {2012})}\BibitemShut {NoStop}%
\bibitem [{\citenamefont {Kim}\ and\ \citenamefont {Sayama}(2017)}]{KIM17}%
  \BibitemOpen
  \bibfield  {author} {\bibinfo {author} {\bibfnamefont {M.}~\bibnamefont
  {Kim}}\ and\ \bibinfo {author} {\bibfnamefont {H.}~\bibnamefont {Sayama}},\
  }\href@noop {} {\bibfield  {journal} {\bibinfo  {journal} {Applied network
  science}\ }\textbf {\bibinfo {volume} {2}},\ \bibinfo {pages} {1} (\bibinfo
  {year} {2017})}\BibitemShut {NoStop}%
\bibitem [{\citenamefont {Barrat}\ \emph {et~al.}(2008)\citenamefont {Barrat},
  \citenamefont {Barthelemy},\ and\ \citenamefont {Vespignani}}]{BAR08}%
  \BibitemOpen
  \bibfield  {author} {\bibinfo {author} {\bibfnamefont {A.}~\bibnamefont
  {Barrat}}, \bibinfo {author} {\bibfnamefont {M.}~\bibnamefont {Barthelemy}},
  \ and\ \bibinfo {author} {\bibfnamefont {A.}~\bibnamefont {Vespignani}},\
  }\href@noop {} {\emph {\bibinfo {title} {Dynamical processes on complex
  networks}}}\ (\bibinfo  {publisher} {Cambridge University Press},\ \bibinfo
  {year} {2008})\BibitemShut {NoStop}%
\bibitem [{\citenamefont {Bianconi}(2018)}]{BIA18}%
  \BibitemOpen
  \bibfield  {author} {\bibinfo {author} {\bibfnamefont {G.}~\bibnamefont
  {Bianconi}},\ }\href@noop {} {\emph {\bibinfo {title} {Multilayer networks:
  structure and function}}}\ (\bibinfo  {publisher} {Oxford University Press},\
  \bibinfo {year} {2018})\BibitemShut {NoStop}%
\bibitem [{\citenamefont {Giusti}\ \emph {et~al.}(2016)\citenamefont {Giusti},
  \citenamefont {Ghrist},\ and\ \citenamefont {Bassett}}]{bassett}%
  \BibitemOpen
  \bibfield  {author} {\bibinfo {author} {\bibfnamefont {C.}~\bibnamefont
  {Giusti}}, \bibinfo {author} {\bibfnamefont {R.}~\bibnamefont {Ghrist}}, \
  and\ \bibinfo {author} {\bibfnamefont {D.~S.}\ \bibnamefont {Bassett}},\
  }\href@noop {} {\bibfield  {journal} {\bibinfo  {journal} {Journal of
  computational neuroscience}\ }\textbf {\bibinfo {volume} {41}},\ \bibinfo
  {pages} {1} (\bibinfo {year} {2016})}\BibitemShut {NoStop}%
\bibitem [{\citenamefont {Bianconi}(2021)}]{bianconi2021higher}%
  \BibitemOpen
  \bibfield  {author} {\bibinfo {author} {\bibfnamefont {G.}~\bibnamefont
  {Bianconi}},\ }\href@noop {} {\emph {\bibinfo {title} {Higher-order networks:
  An introduction to simplicial complexes}}}\ (\bibinfo  {publisher} {Cambridge
  University Press},\ \bibinfo {year} {(in Press) 2021})\BibitemShut {NoStop}%
\bibitem [{\citenamefont {Battiston}\ \emph {et~al.}(2020)\citenamefont
  {Battiston}, \citenamefont {Cencetti}, \citenamefont {Iacopini},
  \citenamefont {Latora}, \citenamefont {Lucas}, \citenamefont {Patania},
  \citenamefont {Young},\ and\ \citenamefont {Petri}}]{battiston}%
  \BibitemOpen
  \bibfield  {author} {\bibinfo {author} {\bibfnamefont {F.}~\bibnamefont
  {Battiston}}, \bibinfo {author} {\bibfnamefont {G.}~\bibnamefont {Cencetti}},
  \bibinfo {author} {\bibfnamefont {I.}~\bibnamefont {Iacopini}}, \bibinfo
  {author} {\bibfnamefont {V.}~\bibnamefont {Latora}}, \bibinfo {author}
  {\bibfnamefont {M.}~\bibnamefont {Lucas}}, \bibinfo {author} {\bibfnamefont
  {A.}~\bibnamefont {Patania}}, \bibinfo {author} {\bibfnamefont {J.-G.}\
  \bibnamefont {Young}}, \ and\ \bibinfo {author} {\bibfnamefont
  {G.}~\bibnamefont {Petri}},\ }\href@noop {} {\bibfield  {journal} {\bibinfo
  {journal} {Physics Reports}\ }\textbf {\bibinfo {volume} {874}},\ \bibinfo
  {pages} {1} (\bibinfo {year} {2020})}\BibitemShut {NoStop}%
\bibitem [{\citenamefont {Torres}\ \emph {et~al.}(2021)\citenamefont {Torres},
  \citenamefont {Blevins}, \citenamefont {Bassett},\ and\ \citenamefont
  {Eliassi-Rad}}]{tina}%
  \BibitemOpen
  \bibfield  {author} {\bibinfo {author} {\bibfnamefont {L.}~\bibnamefont
  {Torres}}, \bibinfo {author} {\bibfnamefont {A.~S.}\ \bibnamefont {Blevins}},
  \bibinfo {author} {\bibfnamefont {D.}~\bibnamefont {Bassett}}, \ and\
  \bibinfo {author} {\bibfnamefont {T.}~\bibnamefont {Eliassi-Rad}},\
  }\href@noop {} {\bibfield  {journal} {\bibinfo  {journal} {SIAM Review}\
  }\textbf {\bibinfo {volume} {63}},\ \bibinfo {pages} {435} (\bibinfo {year}
  {2021})}\BibitemShut {NoStop}%
\bibitem [{\citenamefont {Sizemore}\ \emph {et~al.}(2018)\citenamefont
  {Sizemore}, \citenamefont {Giusti}, \citenamefont {Kahn}, \citenamefont
  {Vettel}, \citenamefont {Betzel},\ and\ \citenamefont {Bassett}}]{SIZ18}%
  \BibitemOpen
  \bibfield  {author} {\bibinfo {author} {\bibfnamefont {A.~E.}\ \bibnamefont
  {Sizemore}}, \bibinfo {author} {\bibfnamefont {C.}~\bibnamefont {Giusti}},
  \bibinfo {author} {\bibfnamefont {A.}~\bibnamefont {Kahn}}, \bibinfo {author}
  {\bibfnamefont {J.~M.}\ \bibnamefont {Vettel}}, \bibinfo {author}
  {\bibfnamefont {R.~F.}\ \bibnamefont {Betzel}}, \ and\ \bibinfo {author}
  {\bibfnamefont {D.~S.}\ \bibnamefont {Bassett}},\ }\href@noop {} {\bibfield
  {journal} {\bibinfo  {journal} {Journal of computational neuroscience}\
  }\textbf {\bibinfo {volume} {44}},\ \bibinfo {pages} {115} (\bibinfo {year}
  {2018})}\BibitemShut {NoStop}%
\bibitem [{\citenamefont {Andjelkovi{\'c}}\ \emph {et~al.}(2020)\citenamefont
  {Andjelkovi{\'c}}, \citenamefont {Tadi{\'c}},\ and\ \citenamefont
  {Melnik}}]{AND20}%
  \BibitemOpen
  \bibfield  {author} {\bibinfo {author} {\bibfnamefont {M.}~\bibnamefont
  {Andjelkovi{\'c}}}, \bibinfo {author} {\bibfnamefont {B.}~\bibnamefont
  {Tadi{\'c}}}, \ and\ \bibinfo {author} {\bibfnamefont {R.}~\bibnamefont
  {Melnik}},\ }\href@noop {} {\bibfield  {journal} {\bibinfo  {journal}
  {Scientific reports}\ }\textbf {\bibinfo {volume} {10}},\ \bibinfo {pages}
  {1} (\bibinfo {year} {2020})}\BibitemShut {NoStop}%
\bibitem [{\citenamefont {Estrada}\ and\ \citenamefont {Ross}(2018)}]{EST18}%
  \BibitemOpen
  \bibfield  {author} {\bibinfo {author} {\bibfnamefont {E.}~\bibnamefont
  {Estrada}}\ and\ \bibinfo {author} {\bibfnamefont {G.~J.}\ \bibnamefont
  {Ross}},\ }\href@noop {} {\bibfield  {journal} {\bibinfo  {journal} {Journal
  of theoretical biology}\ }\textbf {\bibinfo {volume} {438}},\ \bibinfo
  {pages} {46} (\bibinfo {year} {2018})}\BibitemShut {NoStop}%
\bibitem [{\citenamefont {Salnikov}\ \emph {et~al.}(2018)\citenamefont
  {Salnikov}, \citenamefont {Cassese},\ and\ \citenamefont
  {Lambiotte}}]{SAL18}%
  \BibitemOpen
  \bibfield  {author} {\bibinfo {author} {\bibfnamefont {V.}~\bibnamefont
  {Salnikov}}, \bibinfo {author} {\bibfnamefont {D.}~\bibnamefont {Cassese}}, \
  and\ \bibinfo {author} {\bibfnamefont {R.}~\bibnamefont {Lambiotte}},\
  }\href@noop {} {\bibfield  {journal} {\bibinfo  {journal} {European Journal
  of Physics}\ }\textbf {\bibinfo {volume} {40}},\ \bibinfo {pages} {014001}
  (\bibinfo {year} {2018})}\BibitemShut {NoStop}%
\bibitem [{\citenamefont {Benson}\ \emph {et~al.}(2016)\citenamefont {Benson},
  \citenamefont {Gleich},\ and\ \citenamefont {Leskovec}}]{BEN16}%
  \BibitemOpen
  \bibfield  {author} {\bibinfo {author} {\bibfnamefont {A.~R.}\ \bibnamefont
  {Benson}}, \bibinfo {author} {\bibfnamefont {D.~F.}\ \bibnamefont {Gleich}},
  \ and\ \bibinfo {author} {\bibfnamefont {J.}~\bibnamefont {Leskovec}},\
  }\href@noop {} {\bibfield  {journal} {\bibinfo  {journal} {Science}\ }\textbf
  {\bibinfo {volume} {353}},\ \bibinfo {pages} {163} (\bibinfo {year}
  {2016})}\BibitemShut {NoStop}%
\bibitem [{\citenamefont {Schaub}\ \emph {et~al.}(2020)\citenamefont {Schaub},
  \citenamefont {Benson}, \citenamefont {Horn}, \citenamefont {Lippner},\ and\
  \citenamefont {Jadbabaie}}]{SCH20}%
  \BibitemOpen
  \bibfield  {author} {\bibinfo {author} {\bibfnamefont {M.~T.}\ \bibnamefont
  {Schaub}}, \bibinfo {author} {\bibfnamefont {A.~R.}\ \bibnamefont {Benson}},
  \bibinfo {author} {\bibfnamefont {P.}~\bibnamefont {Horn}}, \bibinfo {author}
  {\bibfnamefont {G.}~\bibnamefont {Lippner}}, \ and\ \bibinfo {author}
  {\bibfnamefont {A.}~\bibnamefont {Jadbabaie}},\ }\href@noop {} {\bibfield
  {journal} {\bibinfo  {journal} {SIAM Review}\ }\textbf {\bibinfo {volume}
  {62}},\ \bibinfo {pages} {353} (\bibinfo {year} {2020})}\BibitemShut
  {NoStop}%
\bibitem [{\citenamefont {Iacopini}\ \emph {et~al.}(2019)\citenamefont
  {Iacopini}, \citenamefont {Petri}, \citenamefont {Barrat},\ and\
  \citenamefont {Latora}}]{IAC19}%
  \BibitemOpen
  \bibfield  {author} {\bibinfo {author} {\bibfnamefont {I.}~\bibnamefont
  {Iacopini}}, \bibinfo {author} {\bibfnamefont {G.}~\bibnamefont {Petri}},
  \bibinfo {author} {\bibfnamefont {A.}~\bibnamefont {Barrat}}, \ and\ \bibinfo
  {author} {\bibfnamefont {V.}~\bibnamefont {Latora}},\ }\href@noop {}
  {\bibfield  {journal} {\bibinfo  {journal} {Nature communications}\ }\textbf
  {\bibinfo {volume} {10}},\ \bibinfo {pages} {1} (\bibinfo {year}
  {2019})}\BibitemShut {NoStop}%
\bibitem [{\citenamefont {Estrada}\ and\ \citenamefont
  {Rodriguez-Velazquez}(2005)}]{EST05}%
  \BibitemOpen
  \bibfield  {author} {\bibinfo {author} {\bibfnamefont {E.}~\bibnamefont
  {Estrada}}\ and\ \bibinfo {author} {\bibfnamefont {J.~A.}\ \bibnamefont
  {Rodriguez-Velazquez}},\ }\href@noop {} {\bibfield  {journal} {\bibinfo
  {journal} {arXiv preprint physics/0505137}\ } (\bibinfo {year}
  {2005})}\BibitemShut {NoStop}%
\bibitem [{\citenamefont {Torres}\ and\ \citenamefont
  {Bianconi}(2020)}]{joaquin}%
  \BibitemOpen
  \bibfield  {author} {\bibinfo {author} {\bibfnamefont {J.~J.}\ \bibnamefont
  {Torres}}\ and\ \bibinfo {author} {\bibfnamefont {G.}~\bibnamefont
  {Bianconi}},\ }\href@noop {} {\bibfield  {journal} {\bibinfo  {journal}
  {Journal of Physics: Complexity}\ }\textbf {\bibinfo {volume} {1}},\ \bibinfo
  {pages} {015002} (\bibinfo {year} {2020})}\BibitemShut {NoStop}%
\bibitem [{\citenamefont {Mill{\'a}n}\ \emph {et~al.}(2020)\citenamefont
  {Mill{\'a}n}, \citenamefont {Torres},\ and\ \citenamefont
  {Bianconi}}]{millan}%
  \BibitemOpen
  \bibfield  {author} {\bibinfo {author} {\bibfnamefont {A.~P.}\ \bibnamefont
  {Mill{\'a}n}}, \bibinfo {author} {\bibfnamefont {J.~J.}\ \bibnamefont
  {Torres}}, \ and\ \bibinfo {author} {\bibfnamefont {G.}~\bibnamefont
  {Bianconi}},\ }\href@noop {} {\bibfield  {journal} {\bibinfo  {journal}
  {Physical Review Letters}\ }\textbf {\bibinfo {volume} {124}},\ \bibinfo
  {pages} {218301} (\bibinfo {year} {2020})}\BibitemShut {NoStop}%
\bibitem [{\citenamefont {Ghorbanchian}\ \emph {et~al.}(2021)\citenamefont
  {Ghorbanchian}, \citenamefont {Restrepo}, \citenamefont {Torres},\ and\
  \citenamefont {Bianconi}}]{reza}%
  \BibitemOpen
  \bibfield  {author} {\bibinfo {author} {\bibfnamefont {R.}~\bibnamefont
  {Ghorbanchian}}, \bibinfo {author} {\bibfnamefont {J.~G.}\ \bibnamefont
  {Restrepo}}, \bibinfo {author} {\bibfnamefont {J.~J.}\ \bibnamefont
  {Torres}}, \ and\ \bibinfo {author} {\bibfnamefont {G.}~\bibnamefont
  {Bianconi}},\ }\href@noop {} {\bibfield  {journal} {\bibinfo  {journal}
  {Communications Physics}\ }\textbf {\bibinfo {volume} {4}},\ \bibinfo {pages}
  {1} (\bibinfo {year} {2021})}\BibitemShut {NoStop}%
\bibitem [{\citenamefont {Calmon}\ \emph {et~al.}(2021)\citenamefont {Calmon},
  \citenamefont {Restrepo}, \citenamefont {Torres},\ and\ \citenamefont
  {Bianconi}}]{calmon}%
  \BibitemOpen
  \bibfield  {author} {\bibinfo {author} {\bibfnamefont {L.}~\bibnamefont
  {Calmon}}, \bibinfo {author} {\bibfnamefont {J.~G.}\ \bibnamefont
  {Restrepo}}, \bibinfo {author} {\bibfnamefont {J.~J.}\ \bibnamefont
  {Torres}}, \ and\ \bibinfo {author} {\bibfnamefont {G.}~\bibnamefont
  {Bianconi}},\ }\href@noop {} {\bibfield  {journal} {\bibinfo  {journal}
  {arXiv preprint arXiv:2107.05107}\ } (\bibinfo {year} {2021})}\BibitemShut
  {NoStop}%
\bibitem [{\citenamefont {Jiang}\ \emph {et~al.}(2011)\citenamefont {Jiang},
  \citenamefont {Lim}, \citenamefont {Yao},\ and\ \citenamefont {Ye}}]{JIA11}%
  \BibitemOpen
  \bibfield  {author} {\bibinfo {author} {\bibfnamefont {X.}~\bibnamefont
  {Jiang}}, \bibinfo {author} {\bibfnamefont {L.-H.}\ \bibnamefont {Lim}},
  \bibinfo {author} {\bibfnamefont {Y.}~\bibnamefont {Yao}}, \ and\ \bibinfo
  {author} {\bibfnamefont {Y.}~\bibnamefont {Ye}},\ }\href@noop {} {\bibfield
  {journal} {\bibinfo  {journal} {Mathematical Programming}\ }\textbf {\bibinfo
  {volume} {127}},\ \bibinfo {pages} {203} (\bibinfo {year}
  {2011})}\BibitemShut {NoStop}%
\bibitem [{\citenamefont {Stolz}\ \emph {et~al.}(2017)\citenamefont {Stolz},
  \citenamefont {Harrington},\ and\ \citenamefont {Porter}}]{STO17}%
  \BibitemOpen
  \bibfield  {author} {\bibinfo {author} {\bibfnamefont {B.~J.}\ \bibnamefont
  {Stolz}}, \bibinfo {author} {\bibfnamefont {H.~A.}\ \bibnamefont
  {Harrington}}, \ and\ \bibinfo {author} {\bibfnamefont {M.~A.}\ \bibnamefont
  {Porter}},\ }\href@noop {} {\bibfield  {journal} {\bibinfo  {journal} {Chaos:
  An Interdisciplinary Journal of Nonlinear Science}\ }\textbf {\bibinfo
  {volume} {27}},\ \bibinfo {pages} {047410} (\bibinfo {year}
  {2017})}\BibitemShut {NoStop}%
\bibitem [{\citenamefont {Wu}\ \emph {et~al.}(2015)\citenamefont {Wu},
  \citenamefont {Menichetti}, \citenamefont {Rahmede},\ and\ \citenamefont
  {Bianconi}}]{WU15}%
  \BibitemOpen
  \bibfield  {author} {\bibinfo {author} {\bibfnamefont {Z.}~\bibnamefont
  {Wu}}, \bibinfo {author} {\bibfnamefont {G.}~\bibnamefont {Menichetti}},
  \bibinfo {author} {\bibfnamefont {C.}~\bibnamefont {Rahmede}}, \ and\
  \bibinfo {author} {\bibfnamefont {G.}~\bibnamefont {Bianconi}},\ }\href@noop
  {} {\bibfield  {journal} {\bibinfo  {journal} {Scientific reports}\ }\textbf
  {\bibinfo {volume} {5}},\ \bibinfo {pages} {1} (\bibinfo {year}
  {2015})}\BibitemShut {NoStop}%
\bibitem [{\citenamefont {Bianconi}\ and\ \citenamefont
  {Rahmede}(2016)}]{BIA16}%
  \BibitemOpen
  \bibfield  {author} {\bibinfo {author} {\bibfnamefont {G.}~\bibnamefont
  {Bianconi}}\ and\ \bibinfo {author} {\bibfnamefont {C.}~\bibnamefont
  {Rahmede}},\ }\href@noop {} {\bibfield  {journal} {\bibinfo  {journal}
  {Physical Review E}\ }\textbf {\bibinfo {volume} {93}},\ \bibinfo {pages}
  {032315} (\bibinfo {year} {2016})}\BibitemShut {NoStop}%
\bibitem [{\citenamefont {Bianconi}\ and\ \citenamefont
  {Rahmede}(2017)}]{BIA17}%
  \BibitemOpen
  \bibfield  {author} {\bibinfo {author} {\bibfnamefont {G.}~\bibnamefont
  {Bianconi}}\ and\ \bibinfo {author} {\bibfnamefont {C.}~\bibnamefont
  {Rahmede}},\ }\href@noop {} {\bibfield  {journal} {\bibinfo  {journal}
  {Scientific reports}\ }\textbf {\bibinfo {volume} {7}},\ \bibinfo {pages} {1}
  (\bibinfo {year} {2017})}\BibitemShut {NoStop}%
\bibitem [{\citenamefont {Girvan}\ and\ \citenamefont {Newman}(2002)}]{GIR02}%
  \BibitemOpen
  \bibfield  {author} {\bibinfo {author} {\bibfnamefont {M.}~\bibnamefont
  {Girvan}}\ and\ \bibinfo {author} {\bibfnamefont {M.~E.}\ \bibnamefont
  {Newman}},\ }\href@noop {} {\bibfield  {journal} {\bibinfo  {journal}
  {Proceedings of the national academy of sciences}\ }\textbf {\bibinfo
  {volume} {99}},\ \bibinfo {pages} {7821} (\bibinfo {year}
  {2002})}\BibitemShut {NoStop}%
\bibitem [{\citenamefont {Fortunato}\ and\ \citenamefont {Hric}(2016)}]{FOR16}%
  \BibitemOpen
  \bibfield  {author} {\bibinfo {author} {\bibfnamefont {S.}~\bibnamefont
  {Fortunato}}\ and\ \bibinfo {author} {\bibfnamefont {D.}~\bibnamefont
  {Hric}},\ }\href@noop {} {\bibfield  {journal} {\bibinfo  {journal} {Physics
  reports}\ }\textbf {\bibinfo {volume} {659}},\ \bibinfo {pages} {1} (\bibinfo
  {year} {2016})}\BibitemShut {NoStop}%
\bibitem [{\citenamefont {Blondel}\ \emph {et~al.}(2008)\citenamefont
  {Blondel}, \citenamefont {Guillaume}, \citenamefont {Lambiotte},\ and\
  \citenamefont {Lefebvre}}]{BLO08}%
  \BibitemOpen
  \bibfield  {author} {\bibinfo {author} {\bibfnamefont {V.~D.}\ \bibnamefont
  {Blondel}}, \bibinfo {author} {\bibfnamefont {J.-L.}\ \bibnamefont
  {Guillaume}}, \bibinfo {author} {\bibfnamefont {R.}~\bibnamefont
  {Lambiotte}}, \ and\ \bibinfo {author} {\bibfnamefont {E.}~\bibnamefont
  {Lefebvre}},\ }\href@noop {} {\bibfield  {journal} {\bibinfo  {journal}
  {Journal of statistical mechanics: theory and experiment}\ }\textbf {\bibinfo
  {volume} {2008}},\ \bibinfo {pages} {P10008} (\bibinfo {year}
  {2008})}\BibitemShut {NoStop}%
\bibitem [{\citenamefont {Shai}\ \emph {et~al.}(2017)\citenamefont {Shai},
  \citenamefont {Stanley}, \citenamefont {Granell}, \citenamefont {Taylor},\
  and\ \citenamefont {Mucha}}]{SHA17}%
  \BibitemOpen
  \bibfield  {author} {\bibinfo {author} {\bibfnamefont {S.}~\bibnamefont
  {Shai}}, \bibinfo {author} {\bibfnamefont {N.}~\bibnamefont {Stanley}},
  \bibinfo {author} {\bibfnamefont {C.}~\bibnamefont {Granell}}, \bibinfo
  {author} {\bibfnamefont {D.}~\bibnamefont {Taylor}}, \ and\ \bibinfo {author}
  {\bibfnamefont {P.~J.}\ \bibnamefont {Mucha}},\ }in\ \href@noop {} {\emph
  {\bibinfo {booktitle} {The Oxford Handbook of Social Networks}}}\ (\bibinfo
  {year} {2017})\BibitemShut {NoStop}%
\bibitem [{\citenamefont {Palla}\ \emph
  {et~al.}(2005{\natexlab{a}})\citenamefont {Palla}, \citenamefont
  {Der{\'e}nyi}, \citenamefont {Farkas},\ and\ \citenamefont
  {Vicsek}}]{palla2005uncovering}%
  \BibitemOpen
  \bibfield  {author} {\bibinfo {author} {\bibfnamefont {G.}~\bibnamefont
  {Palla}}, \bibinfo {author} {\bibfnamefont {I.}~\bibnamefont {Der{\'e}nyi}},
  \bibinfo {author} {\bibfnamefont {I.}~\bibnamefont {Farkas}}, \ and\ \bibinfo
  {author} {\bibfnamefont {T.}~\bibnamefont {Vicsek}},\ }\href@noop {}
  {\bibfield  {journal} {\bibinfo  {journal} {nature}\ }\textbf {\bibinfo
  {volume} {435}},\ \bibinfo {pages} {814} (\bibinfo {year}
  {2005}{\natexlab{a}})}\BibitemShut {NoStop}%
\bibitem [{\citenamefont {Betzel}(2020)}]{BET20}%
  \BibitemOpen
  \bibfield  {author} {\bibinfo {author} {\bibfnamefont {R.~F.}\ \bibnamefont
  {Betzel}},\ }\href@noop {} {\bibfield  {journal} {\bibinfo  {journal} {arXiv
  preprint arXiv:2011.06723}\ } (\bibinfo {year} {2020})}\BibitemShut {NoStop}%
\bibitem [{\citenamefont {Van Den~Heuvel}\ and\ \citenamefont
  {Sporns}(2011)}]{VAN11}%
  \BibitemOpen
  \bibfield  {author} {\bibinfo {author} {\bibfnamefont {M.~P.}\ \bibnamefont
  {Van Den~Heuvel}}\ and\ \bibinfo {author} {\bibfnamefont {O.}~\bibnamefont
  {Sporns}},\ }\href@noop {} {\bibfield  {journal} {\bibinfo  {journal}
  {Journal of Neuroscience}\ }\textbf {\bibinfo {volume} {31}},\ \bibinfo
  {pages} {15775} (\bibinfo {year} {2011})}\BibitemShut {NoStop}%
\bibitem [{\citenamefont {Das}\ \emph {et~al.}(2020)\citenamefont {Das},
  \citenamefont {Katyal}, \citenamefont {Verma}, \citenamefont {Dubey},
  \citenamefont {Singh}, \citenamefont {Agarwal}, \citenamefont {Bhaduri},\
  and\ \citenamefont {Ranjan}}]{DAS20}%
  \BibitemOpen
  \bibfield  {author} {\bibinfo {author} {\bibfnamefont {D.}~\bibnamefont
  {Das}}, \bibinfo {author} {\bibfnamefont {Y.}~\bibnamefont {Katyal}},
  \bibinfo {author} {\bibfnamefont {J.}~\bibnamefont {Verma}}, \bibinfo
  {author} {\bibfnamefont {S.}~\bibnamefont {Dubey}}, \bibinfo {author}
  {\bibfnamefont {A.}~\bibnamefont {Singh}}, \bibinfo {author} {\bibfnamefont
  {K.}~\bibnamefont {Agarwal}}, \bibinfo {author} {\bibfnamefont
  {S.}~\bibnamefont {Bhaduri}}, \ and\ \bibinfo {author} {\bibfnamefont
  {R.}~\bibnamefont {Ranjan}},\ }in\ \href@noop {} {\emph {\bibinfo {booktitle}
  {Proceedings of the 1st Workshop on NLP for COVID-19 at ACL 2020}}}\
  (\bibinfo {year} {2020})\BibitemShut {NoStop}%
\bibitem [{\citenamefont {Chen}\ \emph {et~al.}(2015)\citenamefont {Chen},
  \citenamefont {Xie},\ and\ \citenamefont {Zhang}}]{CHE15}%
  \BibitemOpen
  \bibfield  {author} {\bibinfo {author} {\bibfnamefont {Z.}~\bibnamefont
  {Chen}}, \bibinfo {author} {\bibfnamefont {Z.}~\bibnamefont {Xie}}, \ and\
  \bibinfo {author} {\bibfnamefont {Q.}~\bibnamefont {Zhang}},\ }\href@noop {}
  {\bibfield  {journal} {\bibinfo  {journal} {Neurocomputing}\ }\textbf
  {\bibinfo {volume} {170}},\ \bibinfo {pages} {384} (\bibinfo {year}
  {2015})}\BibitemShut {NoStop}%
\bibitem [{\citenamefont {Morarescu}\ and\ \citenamefont
  {Girard}(2010)}]{MOR10}%
  \BibitemOpen
  \bibfield  {author} {\bibinfo {author} {\bibfnamefont {I.-C.}\ \bibnamefont
  {Morarescu}}\ and\ \bibinfo {author} {\bibfnamefont {A.}~\bibnamefont
  {Girard}},\ }\href@noop {} {\bibfield  {journal} {\bibinfo  {journal} {IEEE
  Transactions on Automatic Control}\ }\textbf {\bibinfo {volume} {56}},\
  \bibinfo {pages} {1862} (\bibinfo {year} {2010})}\BibitemShut {NoStop}%
\bibitem [{\citenamefont {Ebli}\ and\ \citenamefont {Spreemann}(2019)}]{EBL19}%
  \BibitemOpen
  \bibfield  {author} {\bibinfo {author} {\bibfnamefont {S.}~\bibnamefont
  {Ebli}}\ and\ \bibinfo {author} {\bibfnamefont {G.}~\bibnamefont
  {Spreemann}},\ }in\ \href@noop {} {\emph {\bibinfo {booktitle} {2019 18th
  IEEE International Conference On Machine Learning And Applications
  (ICMLA)}}}\ (\bibinfo {organization} {IEEE},\ \bibinfo {year} {2019})\ pp.\
  \bibinfo {pages} {1083--1090}\BibitemShut {NoStop}%
\bibitem [{\citenamefont {Der{\'e}nyi}\ \emph {et~al.}(2005)\citenamefont
  {Der{\'e}nyi}, \citenamefont {Palla},\ and\ \citenamefont {Vicsek}}]{DER05}%
  \BibitemOpen
  \bibfield  {author} {\bibinfo {author} {\bibfnamefont {I.}~\bibnamefont
  {Der{\'e}nyi}}, \bibinfo {author} {\bibfnamefont {G.}~\bibnamefont {Palla}},
  \ and\ \bibinfo {author} {\bibfnamefont {T.}~\bibnamefont {Vicsek}},\
  }\href@noop {} {\bibfield  {journal} {\bibinfo  {journal} {Physical Review
  Letters}\ }\textbf {\bibinfo {volume} {94}},\ \bibinfo {pages} {160202}
  (\bibinfo {year} {2005})}\BibitemShut {NoStop}%
\bibitem [{\citenamefont {Billings}\ \emph {et~al.}(2019)\citenamefont
  {Billings}, \citenamefont {Hu}, \citenamefont {Lerda}, \citenamefont
  {Medvedev}, \citenamefont {Mottes}, \citenamefont {Onicas}, \citenamefont
  {Santoro},\ and\ \citenamefont {Petri}}]{BIL19}%
  \BibitemOpen
  \bibfield  {author} {\bibinfo {author} {\bibfnamefont {J.~C.~W.}\
  \bibnamefont {Billings}}, \bibinfo {author} {\bibfnamefont {M.}~\bibnamefont
  {Hu}}, \bibinfo {author} {\bibfnamefont {G.}~\bibnamefont {Lerda}}, \bibinfo
  {author} {\bibfnamefont {A.~N.}\ \bibnamefont {Medvedev}}, \bibinfo {author}
  {\bibfnamefont {F.}~\bibnamefont {Mottes}}, \bibinfo {author} {\bibfnamefont
  {A.}~\bibnamefont {Onicas}}, \bibinfo {author} {\bibfnamefont
  {A.}~\bibnamefont {Santoro}}, \ and\ \bibinfo {author} {\bibfnamefont
  {G.}~\bibnamefont {Petri}},\ }\href@noop {} {\bibfield  {journal} {\bibinfo
  {journal} {arXiv preprint arXiv:1906.09068}\ } (\bibinfo {year}
  {2019})}\BibitemShut {NoStop}%
\bibitem [{\citenamefont {Chodrow}\ \emph {et~al.}(2021)\citenamefont
  {Chodrow}, \citenamefont {Veldt},\ and\ \citenamefont
  {Benson}}]{chodrow2021hypergraph}%
  \BibitemOpen
  \bibfield  {author} {\bibinfo {author} {\bibfnamefont {P.~S.}\ \bibnamefont
  {Chodrow}}, \bibinfo {author} {\bibfnamefont {N.}~\bibnamefont {Veldt}}, \
  and\ \bibinfo {author} {\bibfnamefont {A.~R.}\ \bibnamefont {Benson}},\
  }\href@noop {} {\bibfield  {journal} {\bibinfo  {journal} {arXiv e-prints}\
  ,\ \bibinfo {pages} {arXiv}} (\bibinfo {year} {2021})}\BibitemShut {NoStop}%
\bibitem [{\citenamefont {Eriksson}\ \emph {et~al.}(2021)\citenamefont
  {Eriksson}, \citenamefont {Carletti}, \citenamefont {Lambiotte},
  \citenamefont {Rojas},\ and\ \citenamefont {Rosvall}}]{eriksson2021flow}%
  \BibitemOpen
  \bibfield  {author} {\bibinfo {author} {\bibfnamefont {A.}~\bibnamefont
  {Eriksson}}, \bibinfo {author} {\bibfnamefont {T.}~\bibnamefont {Carletti}},
  \bibinfo {author} {\bibfnamefont {R.}~\bibnamefont {Lambiotte}}, \bibinfo
  {author} {\bibfnamefont {A.}~\bibnamefont {Rojas}}, \ and\ \bibinfo {author}
  {\bibfnamefont {M.}~\bibnamefont {Rosvall}},\ }\href@noop {} {\bibfield
  {journal} {\bibinfo  {journal} {arXiv preprint arXiv:2105.04389}\ } (\bibinfo
  {year} {2021})}\BibitemShut {NoStop}%
\bibitem [{\citenamefont {Carletti}\ \emph {et~al.}(2021)\citenamefont
  {Carletti}, \citenamefont {Fanelli},\ and\ \citenamefont
  {Lambiotte}}]{carletti2021random}%
  \BibitemOpen
  \bibfield  {author} {\bibinfo {author} {\bibfnamefont {T.}~\bibnamefont
  {Carletti}}, \bibinfo {author} {\bibfnamefont {D.}~\bibnamefont {Fanelli}}, \
  and\ \bibinfo {author} {\bibfnamefont {R.}~\bibnamefont {Lambiotte}},\
  }\href@noop {} {\bibfield  {journal} {\bibinfo  {journal} {Journal of
  Physics: Complexity}\ }\textbf {\bibinfo {volume} {2}},\ \bibinfo {pages}
  {015011} (\bibinfo {year} {2021})}\BibitemShut {NoStop}%
\bibitem [{\citenamefont {Capocci}\ \emph {et~al.}(2005)\citenamefont
  {Capocci}, \citenamefont {Servedio}, \citenamefont {Caldarelli},\ and\
  \citenamefont {Colaiori}}]{capocci2005detecting}%
  \BibitemOpen
  \bibfield  {author} {\bibinfo {author} {\bibfnamefont {A.}~\bibnamefont
  {Capocci}}, \bibinfo {author} {\bibfnamefont {V.~D.}\ \bibnamefont
  {Servedio}}, \bibinfo {author} {\bibfnamefont {G.}~\bibnamefont
  {Caldarelli}}, \ and\ \bibinfo {author} {\bibfnamefont {F.}~\bibnamefont
  {Colaiori}},\ }\href@noop {} {\bibfield  {journal} {\bibinfo  {journal}
  {Physica A: Statistical Mechanics and its Applications}\ }\textbf {\bibinfo
  {volume} {352}},\ \bibinfo {pages} {669} (\bibinfo {year}
  {2005})}\BibitemShut {NoStop}%
\bibitem [{\citenamefont {Von~Luxburg}(2007)}]{von2007tutorial}%
  \BibitemOpen
  \bibfield  {author} {\bibinfo {author} {\bibfnamefont {U.}~\bibnamefont
  {Von~Luxburg}},\ }\href@noop {} {\bibfield  {journal} {\bibinfo  {journal}
  {Statistics and computing}\ }\textbf {\bibinfo {volume} {17}},\ \bibinfo
  {pages} {395} (\bibinfo {year} {2007})}\BibitemShut {NoStop}%
\bibitem [{\citenamefont {Newman}(2013)}]{NEW13}%
  \BibitemOpen
  \bibfield  {author} {\bibinfo {author} {\bibfnamefont {M.~E.}\ \bibnamefont
  {Newman}},\ }\href@noop {} {\bibfield  {journal} {\bibinfo  {journal}
  {Physical Review E}\ }\textbf {\bibinfo {volume} {88}},\ \bibinfo {pages}
  {042822} (\bibinfo {year} {2013})}\BibitemShut {NoStop}%
\bibitem [{\citenamefont {Zachary}(1977)}]{ZAC77}%
  \BibitemOpen
  \bibfield  {author} {\bibinfo {author} {\bibfnamefont {W.~W.}\ \bibnamefont
  {Zachary}},\ }\href@noop {} {\bibfield  {journal} {\bibinfo  {journal}
  {Journal of anthropological research}\ }\textbf {\bibinfo {volume} {33}},\
  \bibinfo {pages} {452} (\bibinfo {year} {1977})}\BibitemShut {NoStop}%
\bibitem [{\citenamefont {Young}\ \emph {et~al.}(2021)\citenamefont {Young},
  \citenamefont {Petri},\ and\ \citenamefont {Peixoto}}]{young2021hypergraph}%
  \BibitemOpen
  \bibfield  {author} {\bibinfo {author} {\bibfnamefont {J.-G.}\ \bibnamefont
  {Young}}, \bibinfo {author} {\bibfnamefont {G.}~\bibnamefont {Petri}}, \ and\
  \bibinfo {author} {\bibfnamefont {T.~P.}\ \bibnamefont {Peixoto}},\
  }\href@noop {} {\bibfield  {journal} {\bibinfo  {journal} {Communications
  Physics}\ }\textbf {\bibinfo {volume} {4}},\ \bibinfo {pages} {1} (\bibinfo
  {year} {2021})}\BibitemShut {NoStop}%
\bibitem [{\citenamefont {Musciotto}\ \emph {et~al.}(2021)\citenamefont
  {Musciotto}, \citenamefont {Battiston},\ and\ \citenamefont
  {Mantegna}}]{musciotto2021detecting}%
  \BibitemOpen
  \bibfield  {author} {\bibinfo {author} {\bibfnamefont {F.}~\bibnamefont
  {Musciotto}}, \bibinfo {author} {\bibfnamefont {F.}~\bibnamefont
  {Battiston}}, \ and\ \bibinfo {author} {\bibfnamefont {R.~N.}\ \bibnamefont
  {Mantegna}},\ }\href@noop {} {\bibfield  {journal} {\bibinfo  {journal}
  {arXiv preprint arXiv:2103.16484}\ } (\bibinfo {year} {2021})}\BibitemShut
  {NoStop}%
\bibitem [{\citenamefont {Jost}\ and\ \citenamefont
  {Mulas}(2019)}]{jost2019hypergraph}%
  \BibitemOpen
  \bibfield  {author} {\bibinfo {author} {\bibfnamefont {J.}~\bibnamefont
  {Jost}}\ and\ \bibinfo {author} {\bibfnamefont {R.}~\bibnamefont {Mulas}},\
  }\href@noop {} {\bibfield  {journal} {\bibinfo  {journal} {Advances in
  mathematics}\ }\textbf {\bibinfo {volume} {351}},\ \bibinfo {pages} {870}
  (\bibinfo {year} {2019})}\BibitemShut {NoStop}%
\bibitem [{\citenamefont {Mulas}\ \emph {et~al.}(2020)\citenamefont {Mulas},
  \citenamefont {Kuehn},\ and\ \citenamefont {Jost}}]{mulas2020coupled}%
  \BibitemOpen
  \bibfield  {author} {\bibinfo {author} {\bibfnamefont {R.}~\bibnamefont
  {Mulas}}, \bibinfo {author} {\bibfnamefont {C.}~\bibnamefont {Kuehn}}, \ and\
  \bibinfo {author} {\bibfnamefont {J.}~\bibnamefont {Jost}},\ }\href@noop {}
  {\bibfield  {journal} {\bibinfo  {journal} {Physical Review E}\ }\textbf
  {\bibinfo {volume} {101}},\ \bibinfo {pages} {062313} (\bibinfo {year}
  {2020})}\BibitemShut {NoStop}%
\bibitem [{\citenamefont {Kahle}(2009)}]{kahle2009topology}%
  \BibitemOpen
  \bibfield  {author} {\bibinfo {author} {\bibfnamefont {M.}~\bibnamefont
  {Kahle}},\ }\href@noop {} {\bibfield  {journal} {\bibinfo  {journal}
  {Discrete mathematics}\ }\textbf {\bibinfo {volume} {309}},\ \bibinfo {pages}
  {1658} (\bibinfo {year} {2009})}\BibitemShut {NoStop}%
\bibitem [{\citenamefont {Bianconi}\ and\ \citenamefont
  {Marsili}(2006)}]{clique}%
  \BibitemOpen
  \bibfield  {author} {\bibinfo {author} {\bibfnamefont {G.}~\bibnamefont
  {Bianconi}}\ and\ \bibinfo {author} {\bibfnamefont {M.}~\bibnamefont
  {Marsili}},\ }\href@noop {} {\bibfield  {journal} {\bibinfo  {journal} {EPL
  (Europhysics Letters)}\ }\textbf {\bibinfo {volume} {74}},\ \bibinfo {pages}
  {740} (\bibinfo {year} {2006})}\BibitemShut {NoStop}%
\bibitem [{\citenamefont {Courtney}\ and\ \citenamefont
  {Bianconi}(2016)}]{courtney}%
  \BibitemOpen
  \bibfield  {author} {\bibinfo {author} {\bibfnamefont {O.~T.}\ \bibnamefont
  {Courtney}}\ and\ \bibinfo {author} {\bibfnamefont {G.}~\bibnamefont
  {Bianconi}},\ }\href@noop {} {\bibfield  {journal} {\bibinfo  {journal}
  {Physical Review E}\ }\textbf {\bibinfo {volume} {93}},\ \bibinfo {pages}
  {062311} (\bibinfo {year} {2016})}\BibitemShut {NoStop}%
\bibitem [{\citenamefont {Young}\ \emph {et~al.}(2017)\citenamefont {Young},
  \citenamefont {Petri}, \citenamefont {Vaccarino},\ and\ \citenamefont
  {Patania}}]{YOU17}%
  \BibitemOpen
  \bibfield  {author} {\bibinfo {author} {\bibfnamefont {J.-G.}\ \bibnamefont
  {Young}}, \bibinfo {author} {\bibfnamefont {G.}~\bibnamefont {Petri}},
  \bibinfo {author} {\bibfnamefont {F.}~\bibnamefont {Vaccarino}}, \ and\
  \bibinfo {author} {\bibfnamefont {A.}~\bibnamefont {Patania}},\ }\href@noop
  {} {\bibfield  {journal} {\bibinfo  {journal} {Physical Review E}\ }\textbf
  {\bibinfo {volume} {96}},\ \bibinfo {pages} {032312} (\bibinfo {year}
  {2017})}\BibitemShut {NoStop}%
\bibitem [{\citenamefont {Palla}\ \emph
  {et~al.}(2005{\natexlab{b}})\citenamefont {Palla}, \citenamefont
  {Der{\'e}nyi}, \citenamefont {Farkas},\ and\ \citenamefont {Vicsek}}]{PAL05}%
  \BibitemOpen
  \bibfield  {author} {\bibinfo {author} {\bibfnamefont {G.}~\bibnamefont
  {Palla}}, \bibinfo {author} {\bibfnamefont {I.}~\bibnamefont {Der{\'e}nyi}},
  \bibinfo {author} {\bibfnamefont {I.}~\bibnamefont {Farkas}}, \ and\ \bibinfo
  {author} {\bibfnamefont {T.}~\bibnamefont {Vicsek}},\ }\href@noop {}
  {\bibfield  {journal} {\bibinfo  {journal} {nature}\ }\textbf {\bibinfo
  {volume} {435}},\ \bibinfo {pages} {814} (\bibinfo {year}
  {2005}{\natexlab{b}})}\BibitemShut {NoStop}%
\bibitem [{\citenamefont {Fu}\ \emph {et~al.}(2014)\citenamefont {Fu},
  \citenamefont {Kang}, \citenamefont {Zhicun}, \citenamefont {Lansheng},\ and\
  \citenamefont {Jing}}]{FU14}%
  \BibitemOpen
  \bibfield  {author} {\bibinfo {author} {\bibfnamefont {C.}~\bibnamefont
  {Fu}}, \bibinfo {author} {\bibfnamefont {Z.}~\bibnamefont {Kang}}, \bibinfo
  {author} {\bibfnamefont {F.}~\bibnamefont {Zhicun}}, \bibinfo {author}
  {\bibfnamefont {H.}~\bibnamefont {Lansheng}}, \ and\ \bibinfo {author}
  {\bibfnamefont {C.}~\bibnamefont {Jing}},\ }in\ \href@noop {} {\emph
  {\bibinfo {booktitle} {2014 International Conference on Computer, Information
  and Telecommunication Systems (CITS)}}}\ (\bibinfo {organization} {IEEE},\
  \bibinfo {year} {2014})\ pp.\ \bibinfo {pages} {1--5}\BibitemShut {NoStop}%
\bibitem [{\citenamefont {Hao}\ \emph {et~al.}(2015)\citenamefont {Hao},
  \citenamefont {Min}, \citenamefont {Pei}, \citenamefont {Park},\ and\
  \citenamefont {Yang}}]{HAO15}%
  \BibitemOpen
  \bibfield  {author} {\bibinfo {author} {\bibfnamefont {F.}~\bibnamefont
  {Hao}}, \bibinfo {author} {\bibfnamefont {G.}~\bibnamefont {Min}}, \bibinfo
  {author} {\bibfnamefont {Z.}~\bibnamefont {Pei}}, \bibinfo {author}
  {\bibfnamefont {D.-S.}\ \bibnamefont {Park}}, \ and\ \bibinfo {author}
  {\bibfnamefont {L.~T.}\ \bibnamefont {Yang}},\ }\href@noop {} {\bibfield
  {journal} {\bibinfo  {journal} {IEEE Systems Journal}\ }\textbf {\bibinfo
  {volume} {11}},\ \bibinfo {pages} {250} (\bibinfo {year} {2015})}\BibitemShut
  {NoStop}%
\bibitem [{\citenamefont {Gregori}\ \emph {et~al.}(2012)\citenamefont
  {Gregori}, \citenamefont {Lenzini},\ and\ \citenamefont {Mainardi}}]{GRE12}%
  \BibitemOpen
  \bibfield  {author} {\bibinfo {author} {\bibfnamefont {E.}~\bibnamefont
  {Gregori}}, \bibinfo {author} {\bibfnamefont {L.}~\bibnamefont {Lenzini}}, \
  and\ \bibinfo {author} {\bibfnamefont {S.}~\bibnamefont {Mainardi}},\
  }\href@noop {} {\bibfield  {journal} {\bibinfo  {journal} {IEEE Transactions
  on Parallel and Distributed Systems}\ }\textbf {\bibinfo {volume} {24}},\
  \bibinfo {pages} {1651} (\bibinfo {year} {2012})}\BibitemShut {NoStop}%
\bibitem [{\citenamefont {Chung}\ and\ \citenamefont
  {Graham}(1997)}]{chung1997spectral}%
  \BibitemOpen
  \bibfield  {author} {\bibinfo {author} {\bibfnamefont {F.~R.}\ \bibnamefont
  {Chung}}\ and\ \bibinfo {author} {\bibfnamefont {F.~C.}\ \bibnamefont
  {Graham}},\ }\href@noop {} {\emph {\bibinfo {title} {Spectral graph
  theory}}},\ \bibinfo {number} {92}\ (\bibinfo  {publisher} {American
  Mathematical Soc.},\ \bibinfo {year} {1997})\BibitemShut {NoStop}%
\bibitem [{\citenamefont {Horak}\ and\ \citenamefont
  {Jost}(2013)}]{horak2013spectra}%
  \BibitemOpen
  \bibfield  {author} {\bibinfo {author} {\bibfnamefont {D.}~\bibnamefont
  {Horak}}\ and\ \bibinfo {author} {\bibfnamefont {J.}~\bibnamefont {Jost}},\
  }\href@noop {} {\bibfield  {journal} {\bibinfo  {journal} {Advances in
  Mathematics}\ }\textbf {\bibinfo {volume} {244}},\ \bibinfo {pages} {303}
  (\bibinfo {year} {2013})}\BibitemShut {NoStop}%
\bibitem [{\citenamefont {Lim}(2020)}]{LIM20}%
  \BibitemOpen
  \bibfield  {author} {\bibinfo {author} {\bibfnamefont {L.-H.}\ \bibnamefont
  {Lim}},\ }\href@noop {} {\bibfield  {journal} {\bibinfo  {journal} {Siam
  Review}\ }\textbf {\bibinfo {volume} {62}},\ \bibinfo {pages} {685} (\bibinfo
  {year} {2020})}\BibitemShut {NoStop}%
\bibitem [{\citenamefont {Jia}\ \emph {et~al.}(2019)\citenamefont {Jia},
  \citenamefont {Schaub}, \citenamefont {Segarra},\ and\ \citenamefont
  {Benson}}]{JIA19}%
  \BibitemOpen
  \bibfield  {author} {\bibinfo {author} {\bibfnamefont {J.}~\bibnamefont
  {Jia}}, \bibinfo {author} {\bibfnamefont {M.~T.}\ \bibnamefont {Schaub}},
  \bibinfo {author} {\bibfnamefont {S.}~\bibnamefont {Segarra}}, \ and\
  \bibinfo {author} {\bibfnamefont {A.~R.}\ \bibnamefont {Benson}},\ }in\
  \href@noop {} {\emph {\bibinfo {booktitle} {Proceedings of the 25th ACM
  SIGKDD International Conference on Knowledge Discovery \& Data Mining}}}\
  (\bibinfo {year} {2019})\ pp.\ \bibinfo {pages} {761--771}\BibitemShut
  {NoStop}%
\bibitem [{\citenamefont {Guerra}\ \emph {et~al.}(2021)\citenamefont {Guerra},
  \citenamefont {De~Gregorio}, \citenamefont {Fugacci}, \citenamefont {Petri},\
  and\ \citenamefont {Vaccarino}}]{GUE21}%
  \BibitemOpen
  \bibfield  {author} {\bibinfo {author} {\bibfnamefont {M.}~\bibnamefont
  {Guerra}}, \bibinfo {author} {\bibfnamefont {A.}~\bibnamefont {De~Gregorio}},
  \bibinfo {author} {\bibfnamefont {U.}~\bibnamefont {Fugacci}}, \bibinfo
  {author} {\bibfnamefont {G.}~\bibnamefont {Petri}}, \ and\ \bibinfo {author}
  {\bibfnamefont {F.}~\bibnamefont {Vaccarino}},\ }\href@noop {} {\bibfield
  {journal} {\bibinfo  {journal} {Scientific reports}\ }\textbf {\bibinfo
  {volume} {11}},\ \bibinfo {pages} {1} (\bibinfo {year} {2021})}\BibitemShut
  {NoStop}%
\bibitem [{\citenamefont {Meil{\u{a}}}(2007)}]{MEI98}%
  \BibitemOpen
  \bibfield  {author} {\bibinfo {author} {\bibfnamefont {M.}~\bibnamefont
  {Meil{\u{a}}}},\ }\href@noop {} {\bibfield  {journal} {\bibinfo  {journal}
  {Journal of multivariate analysis}\ }\textbf {\bibinfo {volume} {98}},\
  \bibinfo {pages} {873} (\bibinfo {year} {2007})}\BibitemShut {NoStop}%
\bibitem [{\citenamefont {Shannon}(2001)}]{SHA01}%
  \BibitemOpen
  \bibfield  {author} {\bibinfo {author} {\bibfnamefont {C.~E.}\ \bibnamefont
  {Shannon}},\ }\href@noop {} {\bibfield  {journal} {\bibinfo  {journal} {ACM
  SIGMOBILE mobile computing and communications review}\ }\textbf {\bibinfo
  {volume} {5}},\ \bibinfo {pages} {3} (\bibinfo {year} {2001})}\BibitemShut
  {NoStop}%
\bibitem [{\citenamefont {Newman}(2006)}]{NEW06}%
  \BibitemOpen
  \bibfield  {author} {\bibinfo {author} {\bibfnamefont {M.~E.}\ \bibnamefont
  {Newman}},\ }\href@noop {} {\bibfield  {journal} {\bibinfo  {journal}
  {Physical review E}\ }\textbf {\bibinfo {volume} {74}},\ \bibinfo {pages}
  {036104} (\bibinfo {year} {2006})}\BibitemShut {NoStop}%
\bibitem [{\citenamefont {Petri}\ \emph {et~al.}(2014)\citenamefont {Petri},
  \citenamefont {Expert}, \citenamefont {Turkheimer}, \citenamefont
  {Carhart-Harris}, \citenamefont {Nutt}, \citenamefont {Hellyer},\ and\
  \citenamefont {Vaccarino}}]{petri2014homological}%
  \BibitemOpen
  \bibfield  {author} {\bibinfo {author} {\bibfnamefont {G.}~\bibnamefont
  {Petri}}, \bibinfo {author} {\bibfnamefont {P.}~\bibnamefont {Expert}},
  \bibinfo {author} {\bibfnamefont {F.}~\bibnamefont {Turkheimer}}, \bibinfo
  {author} {\bibfnamefont {R.}~\bibnamefont {Carhart-Harris}}, \bibinfo
  {author} {\bibfnamefont {D.}~\bibnamefont {Nutt}}, \bibinfo {author}
  {\bibfnamefont {P.~J.}\ \bibnamefont {Hellyer}}, \ and\ \bibinfo {author}
  {\bibfnamefont {F.}~\bibnamefont {Vaccarino}},\ }\href@noop {} {\bibfield
  {journal} {\bibinfo  {journal} {Journal of The Royal Society Interface}\
  }\textbf {\bibinfo {volume} {11}},\ \bibinfo {pages} {20140873} (\bibinfo
  {year} {2014})}\BibitemShut {NoStop}%
\bibitem [{\citenamefont {Rieck}\ \emph {et~al.}(2017)\citenamefont {Rieck},
  \citenamefont {Fugacci}, \citenamefont {Lukasczyk},\ and\ \citenamefont
  {Leitte}}]{RIE17}%
  \BibitemOpen
  \bibfield  {author} {\bibinfo {author} {\bibfnamefont {B.}~\bibnamefont
  {Rieck}}, \bibinfo {author} {\bibfnamefont {U.}~\bibnamefont {Fugacci}},
  \bibinfo {author} {\bibfnamefont {J.}~\bibnamefont {Lukasczyk}}, \ and\
  \bibinfo {author} {\bibfnamefont {H.}~\bibnamefont {Leitte}},\ }\href@noop {}
  {\bibfield  {journal} {\bibinfo  {journal} {IEEE transactions on
  visualization and computer graphics}\ }\textbf {\bibinfo {volume} {24}},\
  \bibinfo {pages} {822} (\bibinfo {year} {2017})}\BibitemShut {NoStop}%
\bibitem [{\citenamefont {Vitevitch}\ and\ \citenamefont
  {Castro}(2015)}]{VIT15}%
  \BibitemOpen
  \bibfield  {author} {\bibinfo {author} {\bibfnamefont {M.~S.}\ \bibnamefont
  {Vitevitch}}\ and\ \bibinfo {author} {\bibfnamefont {N.}~\bibnamefont
  {Castro}},\ }\href@noop {} {\bibfield  {journal} {\bibinfo  {journal}
  {International journal of speech-language pathology}\ }\textbf {\bibinfo
  {volume} {17}},\ \bibinfo {pages} {13} (\bibinfo {year} {2015})}\BibitemShut
  {NoStop}%
\bibitem [{\citenamefont {Wilks}\ and\ \citenamefont {Meara}(2002)}]{WIL02}%
  \BibitemOpen
  \bibfield  {author} {\bibinfo {author} {\bibfnamefont {C.}~\bibnamefont
  {Wilks}}\ and\ \bibinfo {author} {\bibfnamefont {P.}~\bibnamefont {Meara}},\
  }\href@noop {} {\bibfield  {journal} {\bibinfo  {journal} {Second Language
  Research}\ }\textbf {\bibinfo {volume} {18}},\ \bibinfo {pages} {303}
  (\bibinfo {year} {2002})}\BibitemShut {NoStop}%
\bibitem [{\citenamefont {Giblin}(2013)}]{GIB13}%
  \BibitemOpen
  \bibfield  {author} {\bibinfo {author} {\bibfnamefont {P.}~\bibnamefont
  {Giblin}},\ }\href@noop {} {\emph {\bibinfo {title} {Graphs, surfaces and
  homology: an introduction to algebraic topology}}}\ (\bibinfo  {publisher}
  {Springer Science \& Business Media},\ \bibinfo {year} {2013})\BibitemShut
  {NoStop}%
\end{thebibliography}%
\appendix

\section{Important Algebraic Topology Concepts}
Given a simplicial complex and a field $\mathbb F$, one can analyze its structure through algebraic topology defining chains and cochains. Here we introduce mathematical concepts from algebraic topology that form the framework for topological simplicial analysis.
\subsection{Orientation}
\label{app:orientation}
In order to introduce the main algebraic topology concept that we will use in this work,  we need to  choose consistent orientations for each simplex. The orientation of a simplex is a choice of the equivalence class of permutations of its vertices, where up to even permutation on the ordering of vertices fall in the same equivalence class. Each simplex has two possible orientations, and if an oriented simplex $\sigma$ has been assigned an orientation, then an odd permutation of its vertices is denoted by $-\sigma$. A $0$-simplex (vertice) can only have one orientation, however when we proceed to higher-order simplices, keeping track or orientation becomes important. 

	\begin{figure}[htbp!]
		\centering
	\includegraphics[width=0.7\linewidth]{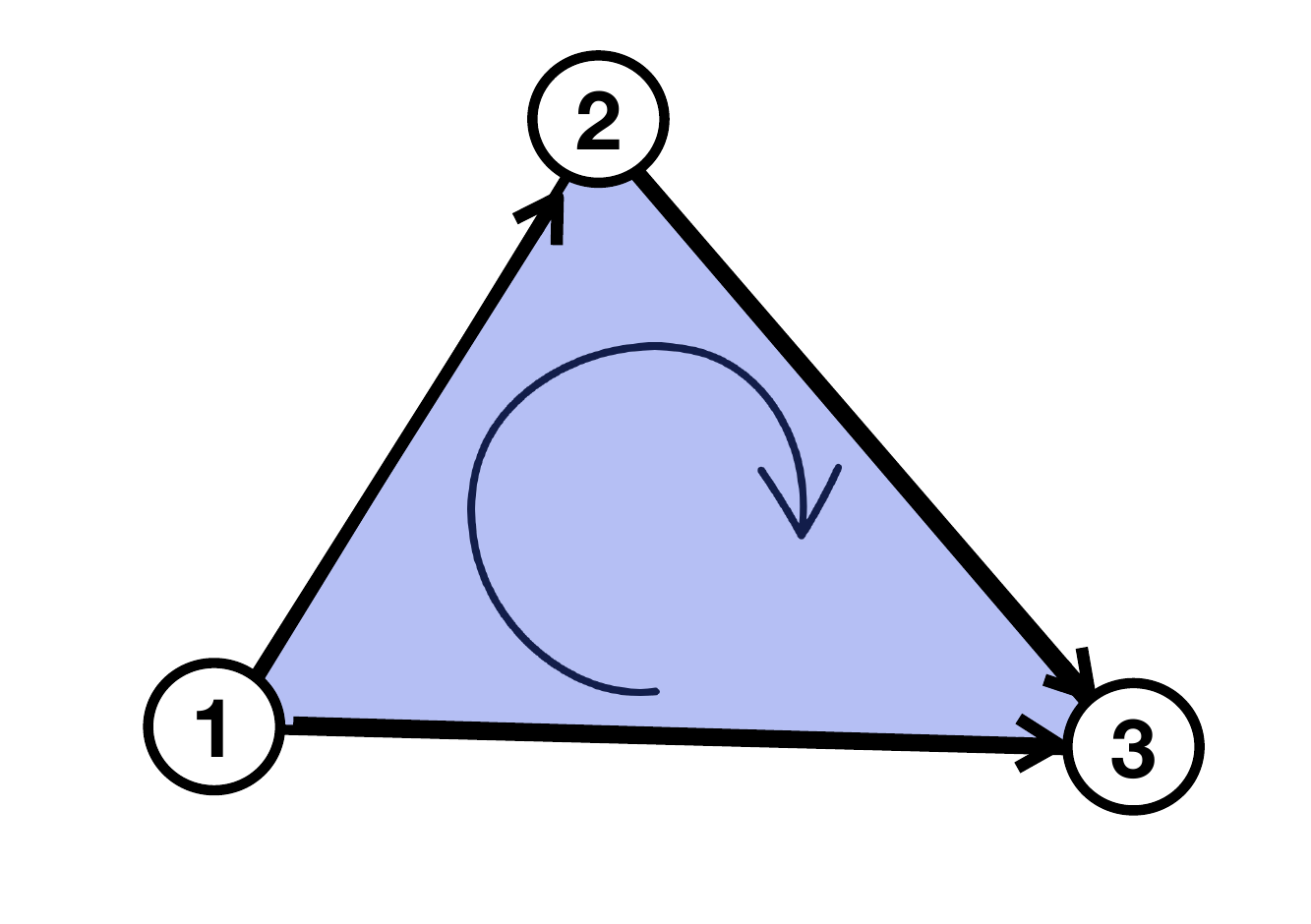}
	\caption{Orientation on simplices determined through labeling the vertices. The triangle is a filled $2$-simplex.}
	\label{orientation}
	\end{figure}

Typically, this is done by assigning an ordering to all vertices in a simplicial complex, and then each simplex inherits an orientation through the induced ordering of its vertices. 
Therefore a simplex $\sigma_k$ given by 
\bea
\sigma_k=[v_0,v_1,\ldots, v_k]
\eea
can be said to have a positive orientation if $v_0<v_1<v_2<\ldots<v_k$ and the simplex obtained from $\sigma_k$ by performing a permutation $\pi$ of the vertices has an orientation determined by the parity of the permutation. For instance, an odd permutation resulting in a simplex defined by $[v_0, v_2, v_1,\ldots v_k]$ involving a single flip is then considered to have a negative orientation.
For instance, in Fig~\ref{orientation}, all  simplices are oriented in the direction of the low vertex-label to the higher vertex label, one can assign this to mean `positive' orientation. Note that orientation, although crucial for bookkeeping is a somewhat artificial concept, and the Hodge Laplacian is orientation independent, as long as the orientation of the simplicial complex is induced by a labeling of the nodes.

\subsection{Chains}
\label{app:boundary}
We define the space $C_k(K)$ of $k$-\emph{chains} on a simplicial complex $K$ as the vector space of \emph{linear combinations of oriented k-simplices}.

In general the fields $\mathbb F = \mathbb R$ and $\mathbb F = \mathbb Z$ are commonly considered. In this work, we consider the field $\mathbb R$.
$C_k$ is a free abelian group, but has the structure of a \emph{vector space} of real functions.
An element ($k$-chain) $c \in C_k$ can be written as a sum of $k$-simplices 
$$
c = \sum_{\sigma\in S_k}w_\sigma \sigma,
$$
where $w_\sigma$ is the weight of each $k$-simplex.
In other words, the $k$-chains are chains of $k$-simplices in a simplicial complex. For instance, in Fig.~\ref{orientation}, $c_1 = [1,2] + [2,3] + [1,3] $ is a $1-$chain. In the real field $\mathbb R$, real weights can be added in front of each $k$-simplex.
\\~\\
Next, we define the \emph{linear boundary maps} between consecutive chain spaces as
$$
\dots \overset{}{\rightarrow} C_{k+1}\overset{\partial_{k+1}}{\rightarrow} C_k\overset{\partial_{k}}{\rightarrow} C_{k-1}\overset{}{\rightarrow}\dots
$$
More precisely, the $k^\text{th}$ \emph{boundary operator} is a linear map $\partial_k:C_k\rightarrow C_{k-1}$ which is determined by its operation on the basis elements of $C_k$ as
$$
\partial_k: \sigma\in C_k \mapsto \sum_{i\in\sigma}(-1)^i(\sigma\backslash \lbrace i\rbrace)
$$
\\
Thus, $\im(\partial_k)$ identifies the image of the operator and is in the space of $(k -1)$-boundaries. Intuitively, the boundary operator acts on a $k$-simplex and returns the $k-1$ simplices that form the faces of the $k$-simplex. For example, the boundary operator applied to the $2-$simplex (filled triangle) in Fig.~\ref{orientation} gives the $1-$simplices that are faces of the triangle, i.e., $\partial_k [1,2,3] = [1,2] + [2,3] - [1,3]$. It is not difficult to show that if we build a cyclic chain $c_k \in C_k$ that starts and ends at the same simplex, then $\partial_k c_k=0$ and vice-versa.  Thus, we call a k-chain $c_k \in \ker(\partial_k)$ a $k$-cycle.

This particular choice of vector spaces $C_k$ and linear operators $d_k$ gives:
\begin{equation}\label{eq: boundary of boundary is zero}
\partial_{k-1}\circ \partial_{k} = 0\text{~for all $k$}.
\end{equation}
In other words, we find that $\im \partial_{k}\subseteq \ker \partial_{k-1}$. 

\subsection{Homology Group}
The boundary operators are linear maps between finite-dimensional vector spaces. After choosing orientations, each of these operators can be represented by a matrix, there by enabling us to perform computations.  We will denote the matrix representation of the boundary operators $\partial_k$ by $B_k$. The $k$-cycles are cycles in the kernel of the boundary operator, i.e., elements in $Z_k := \ker \partial_{k}$ of $\partial_k : C_k \rightarrow C_{k-1}$. 
The $k$-boundaries are cycles that form the boundaries of a $(k+1)$-simplex, i.e., elements in the $B_k : ={\tiny } \im\partial_{k-1}$ of $\partial_{k+1} : C_{k+1} \rightarrow C_{k}$. The cycle $[1,2]+[2,3]-[1,3]$ in Fig. \ref{orientation} is a $1-$boundary since it is the boundary of a 2-simplex. Following expression \eqref{eq: boundary of boundary is zero} we can define the following subspace of $k$-chains 
\begin{equation}
	\label{eq:Hk}
H_k := Z_k/ B_k = \ker \partial_k / \im \partial_{k+1} \text{~i.e. $H_k\subseteq C_k$}.
\end{equation}
The subspace $H_k$ is called the $k^\text{th}$ \emph{homology group} of $K$ \cite{GIB13}, and its dimension $\beta_k:= \dim H_k$ is called the $k^\text{th}$ \emph{Betti number} of $K$. 
These properties are useful because they captures important topological information about the complex. Specifically, the dimension of $H_k$ equals the number of \textit{`$k$-dimensional holes} or cavities in $K$.
\\~\\

\subsection{Cochains and Cospaces}
For completion we also similarly define the cochain vector space denoted by $C^k$ for dimension $k$. They are duals of chains and isomorphic to them, i.e., $C^k (K) := hom C_k(K)$. The basis of these cochains are \textit{functions} on the simplices in the dual chain. Cochains also have the structure of vector spaces, and the coboundary map $\delta_k : C^{k-1} \rightarrow C^k $ can be defined as: 
$$
\delta_k: f (\sigma\in C^k) \mapsto \sum_{i\in\sigma}(-1)^i(f (\sigma) \backslash \lbrace i\rbrace)
$$
\\
the cochain boundaries are linear maps on consecutive cochain spaces as
$$
\dots \overset{}{\leftarrow} C^{k+1}\overset{\delta_{k}}{\leftarrow} C^k\overset{\delta_{k-1}}{\leftarrow} C^{k-1}\overset{}{\leftarrow}\dots
$$
Simply, the coboundary operator can be thought of as a dot product on the k-chains.
\textit{In other words, $\delta_k$ can be viewed as the dual of the boundary map $\partial_{k+1}$}. Additionally $\delta_i\delta_{i-1} =0 $, i.e., the image of $\partial_{i-1}$ is contained in the kernel of $\delta_{i}$. 
The corresponding cohomology group is then given by:
$$
\tilde{H_k} := \ker \delta_k/\im \delta_{k-1}
$$
\\
For each boundary map there exists a coboundary map which is simply its adjoint. The co-boundary operator is denoted is matrix form by $B_k^T$.
 Interesting, one can obtain the graph Laplacian as follows: 
$$
L_0 = B_0 B_0^T
$$
\\

\section{Character Affiliations in Les Mis\'erables Simplicial Communities}
\label{app:lesmiscomm}

The list of $k$-simplicial communities (also called $k$-up communities) in the Les Mis\'erables simplicial complex for varying $k$ are given by: 
1-up communities: 
\begin{itemize} 
	\item Bahorel, Combeferre, Feuilly, Grantaire, Joly, Mabeuf, Marius, Prouvaire \item Blacheville, Combeferre, Dahlia, Fameuil, Fantine, Favourite, Feuilly, Grantaire, Listolier, Mabeuf, Marius, Prouvaire, Tholomyes, Zephine \item Babet, Brujon, Claquesous, Eponine, Gavroche, Gueulemer, Javert, Madame Thenardier, Montparnasse, Valjean \item Babet, Gueulemer, Javert, Valjean \item Gillenormand, Lieutenant Gillenormand, Mademoiselle Baptistine, Mademoiselle Gillenormand, Madame Magloire, Myriel, Valjean \item Babet, Brujon, Claquesous, Courfeyrac, Eponine, Gavroche, Gueulemer, Javert, Mabeuf, Marius, Madame Thenardier, Montparnasse, Thenardier, Valjean \item Bahorel, Combeferre, Feuilly, Grantaire, Joly, Mabeuf, Marius, Madame Hucheloup, Prouvaire \item Bahorel, Bossuet, Combeferre, Courfeyrac, Enjolras, Feuilly, Gavroche, Grantaire, Joly, Mabeuf, Marius, Prouvaire \item Anzelma, Babet, Bahorel, Bamatabois, BaronessT, Bossuet, Brevet, Brujon, Champmathieu, Chenildieu, Claquesous, Cochepaille, Combeferre, Cosette, Courfeyrac, Enjolras, Eponine, Fantine, Fauchelevent, Feuilly, Gavroche, Gillenormand, Grantaire, Gueulemer, Javert, Joly, Judge, Lieutenant Gillenormand, Mabeuf, Marguerite, Marius, Mademoiselle Baptistine, Mademoiselle Gillenormand, Madame Hucheloup, Madame Magloire, Madame Thenardier, Montparnasse, MotherInnocent, Myriel, Perpetue, Pontmercy, Prouvaire, Simplice, Thenardier, Tholomyes, Toussaint, Valjean, Woman1, Woman2 \item Cosette, Gillenormand, Lieutenant Gillenormand, Mademoiselle Gillenormand, Valjean \item Bamatabois, Brevet, Champmathieu, Chenildieu, Cochepaille, Judge, Valjean \item Child1, Child2, Cosette, Gavroche, Javert, Toussaint, Valjean, Woman2 \item Anzelma, Babet, Bahorel, Bamatabois, BaronessT, Bossuet, Brevet, Brujon, Champmathieu, Chenildieu, Child1, Child2, Claquesous, Cochepaille, Combeferre, Cosette, Courfeyrac, Enjolras, Eponine, Fantine, Fauchelevent, Feuilly, Gavroche, Gillenormand, Grantaire, Gueulemer, Javert, Joly, Judge, Lieutenant Gillenormand, Mabeuf, Marguerite, Marius, Mademoiselle Gillenormand, Madame Hucheloup, Madame Thenardier, Montparnasse, MotherInnocent, Perpetue, Pontmercy, Prouvaire, Simplice, Thenardier, Tholomyes, Toussaint, Valjean, Woman1, Woman2 \item Anzelma, Babet, Bahorel, Bamatabois, BaronessT, Bossuet, Brevet, Brujon, Champmathieu, Chenildieu, Claquesous, Cochepaille, Combeferre, Cosette, Courfeyrac, Enjolras, Eponine, Fantine, Fauchelevent, Feuilly, Gavroche, Gillenormand, Grantaire, Gueulemer, Javert, Joly, Judge, Lieutenant Gillenormand, Mabeuf, Marguerite, Marius, Mademoiselle Gillenormand, Madame Hucheloup, Madame Thenardier, Montparnasse, MotherInnocent, Perpetue, Pontmercy, Prouvaire, Simplice, Thenardier, Tholomyes, Toussaint, Valjean, Woman1, Woman2 
\end{itemize} 

2-up communities: \begin{itemize}
	\item Bahorel, Blacheville, Bossuet, Combeferre, Courfeyrac, Dahlia, Enjolras, Fameuil, Fantine, Favourite, Feuilly, Gavroche, Grantaire, Joly, Listolier, Mabeuf, Marius, Prouvaire, Tholomyes, Zephine \item Babet, Bahorel, Bossuet, Brujon, Claquesous, Combeferre, Courfeyrac, Enjolras, Eponine, Feuilly, Gavroche, Gueulemer, Javert, Joly, Mabeuf, Marius, Madame Thenardier, Montparnasse, Thenardier, Valjean \item Bahorel, Bossuet, Combeferre, Courfeyrac, Enjolras, Feuilly, Gavroche, Grantaire, Joly, Mabeuf, Marius, Madame Hucheloup, Prouvaire \item Babet, Brujon, Claquesous, Eponine, Gavroche, Gueulemer, Javert, Madame Thenardier, Montparnasse, Valjean \item Babet, Bahorel, Bamatabois, Bossuet, Brujon, Claquesous, Combeferre, Cosette, Courfeyrac, Enjolras, Eponine, Fantine, Feuilly, Gavroche, Gillenormand, Grantaire, Gueulemer, Javert, Joly, Lieutenant Gillenormand, Mabeuf, Marius, Mademoiselle Gillenormand, Madame Hucheloup, Madame Thenardier, Montparnasse, Prouvaire, Simplice, Thenardier, Toussaint, Valjean, Woman2 \item Babet, Bamatabois, Brevet, Brujon, Champmathieu, Chenildieu, Claquesous, Cochepaille, Eponine, Gavroche, Gueulemer, Javert, Judge, Madame Thenardier, Montparnasse, Valjean \item Bahorel, Bamatabois, Bossuet, Combeferre, Cosette, Courfeyrac, Enjolras, Eponine, Fantine, Feuilly, Gavroche, Javert, Joly, Mabeuf, Marius, Simplice, Toussaint, Valjean, Woman2 \item Cosette, Fantine, Gillenormand, Javert, Lieutenant Gillenormand, Marius, Mademoiselle Gillenormand, Madame Thenardier, Thenardier, Valjean \item Bahorel, Bossuet, Combeferre, Courfeyrac, Enjolras, Feuilly, Gavroche, Grantaire, Joly, Mabeuf, Marius, Prouvaire \item Bahorel, Bossuet, Combeferre, Courfeyrac, Enjolras, Feuilly, Grantaire, Joly, Mabeuf, Marius, Madame Hucheloup, Prouvaire, Valjean \item Mademoiselle Baptistine, Madame Magloire, Myriel, Valjean \item Babet, Claquesous, Gavroche, Gueulemer, Javert, Madame Thenardier, Montparnasse, Valjean \item Anzelma, Babet, Bahorel, Bamatabois, Bossuet, Brujon, Claquesous, Combeferre, Cosette, Courfeyrac, Enjolras, Eponine, Fantine, Feuilly, Gavroche, Gillenormand, Grantaire, Gueulemer, Javert, Joly, Lieutenant Gillenormand, Mabeuf, Marius, Mademoiselle Gillenormand, Madame Hucheloup, Madame Thenardier, Montparnasse, Prouvaire, Simplice, Thenardier, Toussaint, Valjean, Woman2 \item Babet, Brujon, Claquesous, Eponine, Gavroche, Gueulemer, Javert, Madame Thenardier, Montparnasse, Thenardier, Valjean \item Cosette, Gillenormand, Lieutenant Gillenormand, Marius, Mademoiselle Gillenormand, Valjean \item Babet, Claquesous, Gavroche, Gueulemer, Javert, Madame Thenardier, Montparnasse, Thenardier, Valjean \end{itemize} 

3-up communities: \begin{itemize} 
	\item Bahorel, Blacheville, Bossuet, Combeferre, Courfeyrac, Dahlia, Enjolras, Fameuil, Fantine, Favourite, Feuilly, Gavroche, Grantaire, Joly, Listolier, Mabeuf, Marius, Prouvaire, Tholomyes, Zephine \item Babet, Claquesous, Cosette, Fantine, Gavroche, Gueulemer, Javert, Madame Thenardier, Montparnasse, Thenardier, Valjean \item Bahorel, Bossuet, Combeferre, Courfeyrac, Enjolras, Feuilly, Gavroche, Grantaire, Joly, Mabeuf, Marius, Madame Hucheloup, Prouvaire, Valjean \item Bahorel, Bossuet, Combeferre, Courfeyrac, Enjolras, Feuilly, Gavroche, Grantaire, Joly, Mabeuf, Marius, Madame Hucheloup, Prouvaire \item Blacheville, Dahlia, Fameuil, Fantine, Favourite, Listolier, Tholomyes, Zephine \item Bahorel, Bossuet, Combeferre, Courfeyrac, Enjolras, Feuilly, Gavroche, Grantaire, Joly, Mabeuf, Marius, Prouvaire \item Bamatabois, Brevet, Champmathieu, Chenildieu, Cochepaille, Judge, Valjean \item Cosette, Fantine, Javert, Madame Thenardier, Thenardier, Valjean \item Bahorel, Blacheville, Combeferre, Courfeyrac, Dahlia, Enjolras, Fameuil, Fantine, Favourite, Feuilly, Gavroche, Grantaire, Joly, Listolier, Mabeuf, Marius, Prouvaire, Tholomyes, Zephine \item Babet, Brujon, Claquesous, Eponine, Gavroche, Gueulemer, Javert, Madame Thenardier, Montparnasse, Thenardier, Valjean \item Cosette, Gillenormand, Lieutenant Gillenormand, Marius, Mademoiselle Gillenormand, Valjean \item Babet, Claquesous, Gavroche, Gueulemer, Javert, Madame Thenardier, Montparnasse, Thenardier, Valjean \item Bahorel, Blacheville, Bossuet, Combeferre, Dahlia, Enjolras, Fameuil, Fantine, Favourite, Feuilly, Gavroche, Grantaire, Joly, Listolier, Mabeuf, Marius, Prouvaire, Tholomyes, Zephine \end{itemize} 

4-up communities: \begin{itemize} 
	\item Bahorel, Bossuet, Combeferre, Courfeyrac, Enjolras, Feuilly, Gavroche, Grantaire, Joly, Mabeuf, Marius, Madame Hucheloup, Prouvaire \item Blacheville, Dahlia, Fameuil, Fantine, Favourite, Listolier, Tholomyes, Zephine \item Bahorel, Bossuet, Combeferre, Courfeyrac, Enjolras, Feuilly, Gavroche, Grantaire, Joly, Mabeuf, Marius, Prouvaire \item Babet, Brujon, Claquesous, Eponine, Gueulemer, Javert, Madame Thenardier, Montparnasse, Thenardier, Valjean \item Bamatabois, Brevet, Champmathieu, Chenildieu, Cochepaille, Judge, Valjean \item Babet, Brujon, Claquesous, Eponine, Gavroche, Gueulemer, Javert, Madame Thenardier, Montparnasse, Thenardier, Valjean \item Babet, Claquesous, Gavroche, Gueulemer, Javert, Madame Thenardier, Montparnasse, Thenardier, Valjean \end{itemize} 

5-up communities: \begin{itemize}
	\item Bahorel, Blacheville, Bossuet, Combeferre, Courfeyrac, Dahlia, Enjolras, Fameuil, Fantine, Favourite, Feuilly, Gavroche, Grantaire, Joly, Listolier, Mabeuf, Marius, Prouvaire, Tholomyes, Zephine \item Bahorel, Bossuet, Combeferre, Courfeyrac, Enjolras, Feuilly, Gavroche, Grantaire, Joly, Mabeuf, Marius, Madame Hucheloup, Prouvaire \item Blacheville, Dahlia, Fameuil, Fantine, Favourite, Listolier, Tholomyes, Zephine \item Babet, Brujon, Claquesous, Eponine, Gueulemer, Montparnasse, Thenardier \item Bahorel, Bossuet, Combeferre, Courfeyrac, Enjolras, Feuilly, Gavroche, Grantaire, Joly, Mabeuf, Marius, Prouvaire \item Bamatabois, Brevet, Champmathieu, Chenildieu, Cochepaille, Judge, Valjean \item Babet, Brujon, Claquesous, Eponine, Gavroche, Gueulemer, Javert, Madame Thenardier, Montparnasse, Thenardier, Valjean \item Babet, Claquesous, Gavroche, Gueulemer, Javert, Madame Thenardier, Montparnasse, Thenardier, Valjean \end{itemize} 

6-up communities: \begin{itemize} 
	\item Bahorel, Bossuet, Combeferre, Courfeyrac, Enjolras, Feuilly, Gavroche, Grantaire, Joly, Prouvaire \item Blacheville, Dahlia, Fameuil, Fantine, Favourite, Listolier, Tholomyes, Zephine \item Bahorel, Bossuet, Combeferre, Courfeyrac, Enjolras, Feuilly, Gavroche, Grantaire, Joly, Mabeuf, Marius, Madame Hucheloup, Prouvaire \item Bahorel, Bossuet, Combeferre, Courfeyrac, Enjolras, Feuilly, Gavroche, Grantaire, Joly, Mabeuf, Marius, Prouvaire \end{itemize} 

Note that a single character (node) can typically be a part of more than one community. The general structure of these communities agrees with those in the novel.

\end{document}